\documentclass[table,twocolumn]{aastex62}
\usepackage{amstext}
\usepackage{amsmath}
\usepackage{apjfonts}
\usepackage[capbesideposition={top}]{floatrow}
\usepackage{float}

\newcommand{\beqar}{\begin{eqnarray}}
\newcommand{\eeqar}{\end{eqnarray}}

\newcommand{\machw}{\mathcal{M}_{\rm w}}
\newcommand{\rw}{R_{\rm w}}
\newcommand{\rb}{R_{\rm b}}
\newcommand{\rH}{R_{\rm H}}
\newcommand{\cs}{c_{\rm s}}

\newcommand{\msmbh}{M_{\rm SMBH}}
\newcommand{\cinf}{c_\infty}
\newcommand{\rhoinf}{\rho_\infty}
\newcommand{\rbw}{R_{\rm bw}}

\newcommand{\beq}{\begin{equation}}
\newcommand{\eeq}{\end{equation}}

\definecolor{nick}{HTML}{006400}
\newcommand*\nko[1]{\normalfont{\color{black}{#1}}}

\begin{document}
\title{The hydrodynamic evolution of binary black holes embedded within the vertically stratified disks of active galactic nuclei}

\correspondingauthor{Nicholas Kaaz}
\email{nkaaz@u.northwestern.edu}

\author[0000-0002-5375-8232]{Nicholas Kaaz}
\affiliation{Department of Physics \& Astronomy, Northwestern University, Evanston, IL 60202, USA}
\affiliation{Center for Interdisciplinary Exploration \& Research in Astrophysics (CIERA), Evanston, IL 60202, USA}

\author[0000-0003-1735-8263]{Sophie Lund Schr{\o}der}
\affiliation{Niels Bohr Institute, University of Copenhagen,Blegdamsvej 17, DK-2100 Copenhagen, Denmark}

\author[0000-0001-5261-3923]{Jeff J. Andrews}
\affiliation{Center for Interdisciplinary Exploration \& Research in Astrophysics (CIERA), Evanston, IL 60202, USA}

\author[0000-0003-3062-4773]{Andrea Antoni}
\affiliation{Department of Astronomy, University of California, Berkeley, CA 94720, USA}

\author[0000-0003-2558-3102]{Enrico Ramirez-Ruiz}
\affiliation{Department of Astronomy \& Astrophysics, University of California, Santa Cruz, CA 95064, USA}
\affiliation{Niels Bohr Institute, University of Copenhagen,Blegdamsvej 17, DK-2100 Copenhagen, Denmark}

\begin{abstract} 
Stellar-mass black holes can become embedded within the disks of active galactic nuclei (AGNs). Afterwards, their interactions are mediated by their gaseous surroundings. Here, we study the evolution of stellar-mass binary black holes (BBHs) embedded within AGN disks using three-dimensional hydrodynamic simulations and analytic methods, focusing on environments where the AGN disk scale height $H$ is $\gtrsim$ the BBH sphere of influence. We model the local surroundings of the embedded BBHs using a wind tunnel formalism and characterize different accretion regimes based on the local properties of the disk. We develop prescriptions for accretion and drag for embedded BBHs. Using these prescriptions with AGN disk models that can represent the Toomre-unstable outer regions of AGN disks, we study the long-term evolution of BBHs as they migrate through the disk. We find that BBHs typically merge within $\lesssim 1-30\,{\rm Myr}$, increasing their mass significantly in the process, allowing BBHs to enter (or cross) the pair-instability supernova mass gap. The BBH accretion rate often exceeds the Eddington limit, sometimes by several orders of magnitude. Many embedded BBHs will merge before migrating significantly in the disk. We also discuss possible electromagnetic signatures during and following the inspiral, finding that it is generally unlikely for the bolometric luminosity of the BBH to exceed the AGN luminosity. 
\end{abstract}

\section{Introduction}
\label{sec:intro}

Direct observations of Sagittarius A$^*$, the supermassive black hole (SMBH) at the center of our galaxy, indicate the existence of a population of massive stars within its sphere of influence \citep{schodel_2002,ghez_2003,paumard_2006,lu_2009,bartko_2009,bartko_2010}. Furthermore, the growing observational sample of tidal disruption events suggests that relatively young stellar populations are ubiquitous around SMBHs \citep{jamie_2017, french_2020, brenna_2021}. While most active galactic nuclei (AGN) harbor SMBHs with low accretion rates \citep{ptak_2001}, a subset of AGN are commonly believed to harbor accretion disks that are abundant in cold, dense gas \citep{2014ARA&A..52..589H}.
These disks remain `canonically' thin \citep{ss1973} up to scales of $10^{-2}-1$\,{\rm pc} \citep{sirko_goodman_2003} depending on the mass of the SMBH, and at larger radii gradually transition between the gas-dominated accretion disk and the star-dominated galactic disk.

It has recently been realized that a subset of the stars around AGN comprise massive stellar binaries that evolve into binary BHs (BBHs). Over the past several years LIGO/Virgo has detected dozens of BBHs mergers \citep{LIGO_2019a,LIGO_2020a,LIGO_2020b} which are typically attributed to formation through isolated binary evolution \citep{belczynski_2016,giacobbo_2018,kruckow_2018,sophie_2018, bavera_2020}
or dynamical interactions in dense stellar systems \citep{oleary_2009,antonini_perets_2012,samsing_2014,antonini_2017,askar_2017,2017ApJ...840L..14S,banerjee_2018,fragione_2018,rodriguez_2018,samsing_2018,kyle_2018,kyle_2019,dicarlo_2019}. 
However, a potentially significant fraction of all LIGO/Virgo events may be due to BBHs that merge within an AGN disk \citep{mckernan_2012, mckernan_2014, bartos_2017, stone_2017, tagawa_2019}.
Due to the high gas densities within the disk, accretion and gas drag will help the binary inspiral much faster than via three-body hardening and gravitational wave emission alone \citep{andrea_2019}.

The possibility that BBHs may merge within an AGN disk provides the tantalizing prospect that some BBH mergers may produce an associated electromagnetic (EM) counterpart; at the moment of merger, the sudden mass loss and recoil of the product black hole (BH) may shock-heat the surrounding accretion flow, resulting in an optical/UV flare \citep{lippai_2008, corrales_2010, demink_2017, mckernan_2019}. While largely unproven, this scenario was reinvigorated with the recently claimed association of the BBH merger GW190521 with an AGN flare \citep{graham_2020}, although we note that the significance of this association has been questioned \citep{ashton_2020}. This scenario is particularly intriguing because the BHs that merged to generate GW190521 are sufficiently massive that they fall within the pair-instability supernova (PISN) mass gap \citep{GW190521}, suggesting that GW190521 could not have been formed through standard binary evolution scenarios \citep{2020ApJ...894..129S,mohammad_2020,2021ApJ...907L..19V} unless our understanding of PISN is in significant error \citep[e.g.,][]{belczynski_2020}. It is therefore worth exploring the possibility that certain BBH mergers, such as GW190521, may have been formed in an AGN disk. 

Stellar-mass BBHs can end up orbiting in the midplane of an AGN disk via two methods. First, they can form \textit{in situ}: in the outer reaches of AGN disks, the gas becomes Toomre unstable, forming generations of stars already embedded within the disk \citep{toomre_1964,goodman_2003}. The BHs that are birthed by this generation of stars couple to the surrounding gaseous disk and stellar population, causing them to inspiral towards the central SMBH - and, in the case of binary systems, cause their orbital separation to shrink \citep[e.g.,][]{stone_2017}. Second, black holes can form in the surrounding nuclear cluster then migrate into the AGN disk via dynamical interactions: the density of the nuclear cluster is sufficiently high that its most massive constituents - the black holes - will mass segregate into the central regions, nearest the SMBH \citep{morris_1993,rasio_2004}. Here, the gravitational influence of the AGN disk is strong, and the mass-segregated black holes will gradually have their inclinations and eccentricities damped until they occupy circular orbits embedded within the AGN disk \citep{ward_1994, tanaka_ward_2004}. Single, rather than binary, BHs may also become embedded by these same mechanisms and then efficiently pair up via gas-mediated single-single encounters \citep{tagawa_2019}.

Once embedded within the AGN disk, BBH evolution is driven by a combination of three-body and hydrodynamic interactions. Repeated single-double encounters increasingly harden binaries. It is estimated that within $\lesssim10$ encounters, a BBH can be sufficiently hardened such that gravitational waves can merge the binary within a Hubble time \citep{leigh_2018}. At the same time, gaseous torques from the disk influence the center-of-mass motion of the BBH. Depending on the assumed AGN disk profile, these gaseous torques can be negative far from the SMBH and positive close to the SMBH, causing embedded objects to become stuck in `migration traps' \citep{lyra_2010,paardekooper_2010,paardekooper_2011}. Within a populated migration trap, the rate of BH and BBH interactions are increased, potentially leading to rapid merger times \citep{secunda_2019,secunda_2020}. 

In these models, a critical ingredient is gas drag that exerts negative torques on the binary that help it inspiral on short timescales. Typically, this effect is included using semi-analytic prescriptions \citep{tagawa_2016,tagawa_2018,mckernan_2018}. These prescriptions are often motivated by more detailed hydrodynamic simulations of accreting binaries \citep[e.g.,][]{andrea_2019}, with hydrodynamic studies of embedded BBHs being only recently studied. \nko{These studies include two-dimensional global simulations \citep{baruteau_2011,li_2021}, and local shearing box simulations in both two \citep{li_2022a,li_lai_2022a} and three \citep{dempsey_2022} dimensions. These simulations have generally focused on cases where the Hill radius of the embedded binary is larger than the scale height of the AGN disk, though often the scale height can be comparable or larger than the BBH sphere of influence. }

In this work, we use three-dimensional hydrodynamic simulations to study the properties of the accretion flow surrounding embedded BBHs in vertically stratified AGN disks and explore its consequences on their long-term evolution. In Section \ref{sec:approach}, we consider the flow geometry surrounding embedded BBHs and contextualize them within AGN disk models. We describe our numerical method and present the results of our hydrodynamic simulations in Section \ref{sec:results}. We apply these results in Section \ref{sec:evolution}, where we study the long-term evolutionary tracks of embedded BBHs as they migrate through the host disk. Finally, in Section \ref{sec:discussion} we discuss the implications of our results for gravitational-wave sources, possible electromagnetic signatures, potential caveats and then provide a summary of our findings.

\section{Setting the stage}
\label{sec:approach}
\subsection{Characteristic scales}

\begin{figure*}[bht]
     \includegraphics[width=0.7\textwidth]{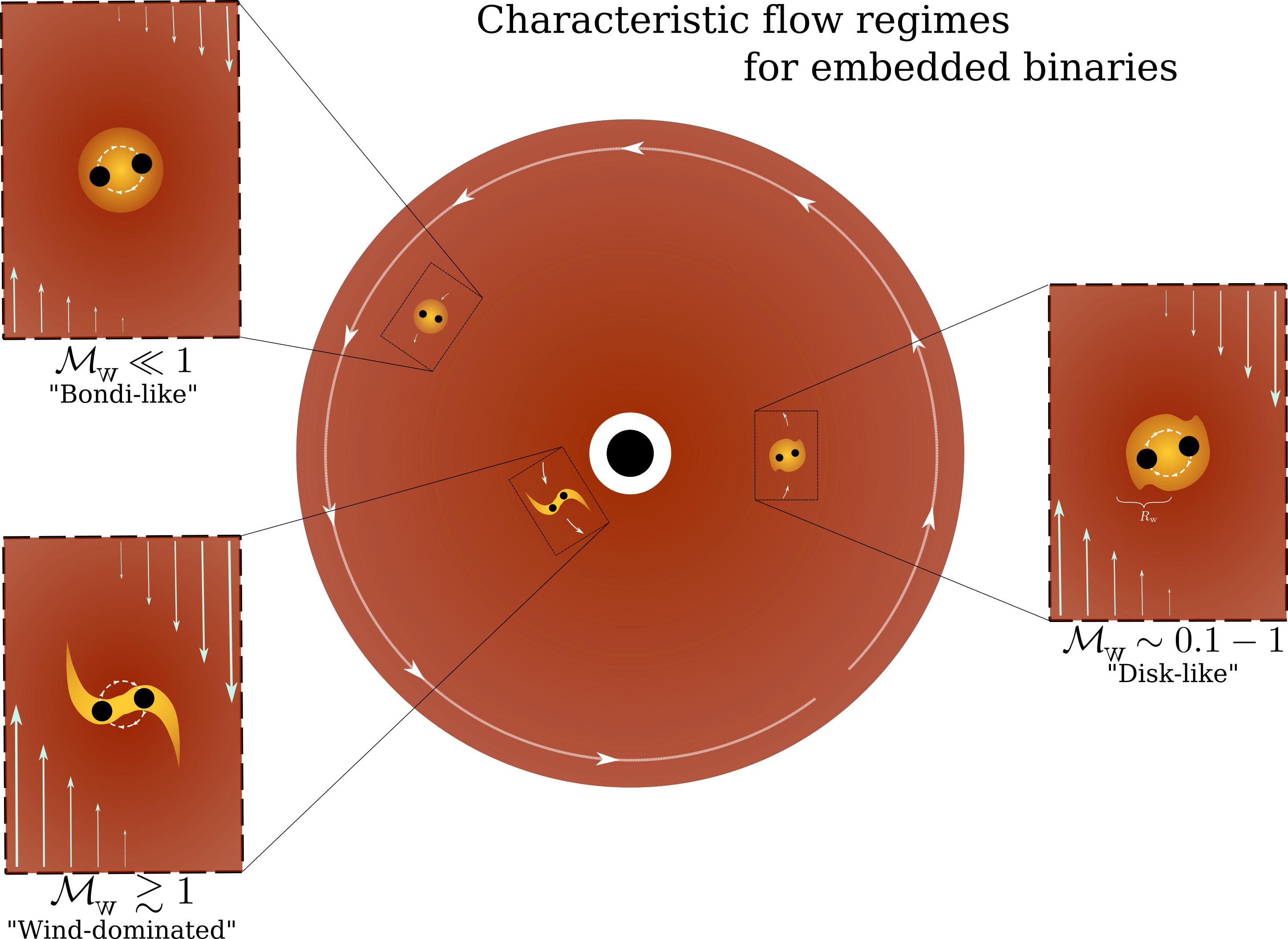}
    \caption{Here, we present a cartoon depiction of the three regimes that we use to describe embedded black hole binaries. We remind the reader that this framework is valid for vertically stratified, three-dimensional accretion flows, but at large $\machw$, $H$ becomes much smaller than the BBH sphere of influence, and the flow geometry becomes planar. The rectangular inset drawings represent the ``wind tunnel'' computational domain that we use in our simulations. \textbf{Upper left:} Far from the central SMBH, the binary sees little of the disk velocity gradient, and accretes in a quasi-spherical Bondi-like fashion. \textbf{Lower left:} Binaries that live near their host SMBH struggle to accrete from the high-velocity disk wind, and gas pressure is unable to thermalize incident streamlines. \textbf{Center right:} For binaries at intermediate distances from the SMBH, the AGN disk deposits angular momentum into the flow, creating structures with partial centrifugal support.}
    \label{fig:cartoon}
\end{figure*}

Consider a BBH that is embedded within an AGN disk. Assume the center of mass of the binary occupies a circular orbit at distance $D$ from the SMBH and assume the center of mass corotates with the disk. Disk annuli closer to the SMBH will orbit faster than the BBH, and annuli farther from the SMBH will orbit slower. In the rest-frame of the BBH, this results in a shearing wind that engulfs the binary. Within some sphere of influence, the BBH will feed from the AGN disk, accreting both mass and angular momentum. The two relevant length scales are the Hill radius of the BBH,
\begin{equation}
    \rH = D\left(\frac{q}{3}\right)^{1/3},
    \label{eq:hill_radius}
\end{equation}
where D is the distance from the SMBH to the BBH center of mass, and $q = M/\msmbh$ is the BBH-SMBH mass ratio. The second length scale to consider is the Bondi radius,
\begin{equation}
    \rb = \frac{GM}{\cs^2},
    \label{eq:bondi_radius}
\end{equation}
were $\cs$ is the local sound speed at distance $D$ in the disk. At radii larger than $\rb$, the pressure support of the gas allows it to be unperturbed by the gravity of the BBH. The sphere of influence of the BBH will be limited by the smaller of these two length-scales. If $\rH < \rb$, the ram pressure of the wind is more important than the gas pressure, and if $\rb < \rH$ the opposite is true. \nko{To parameterize the relative strength of these two regimes, we start by writing down the velocity profile of the wind in the non-inertial rest frame of the BBH,}
\begin{equation}
v(\delta r) = v_{\rm k}(D+\delta r) - v_{\rm k}(D),
\end{equation}
where $v_{\rm k}$ is the Keplerian velocity and $\delta r$ is the relative distance from the BBH. \nko{Here, we have essentially assumed that near ($\pm \delta r$) the BBH, the velocity profile has a Cartesian geometry. We also neglect the vertical velocity dependence of the flow.} It's advantageous to linearize this velocity profile because it will later allow us to model embedded BBHs in a scale-free fashion,
\nko{
\begin{equation}
\begin{aligned}
    v(\delta r)& = \sqrt{\frac{GM_{\rm SMBH}}{D+\delta r}} - \sqrt{\frac{GM_{\rm SMBH}}{D}}\\&\approx \sqrt{\frac{GM_{\rm SMBH}}{D}}\left(1-\frac{\delta r}{2D}\right) - \sqrt{\frac{GM_{\rm SMBH}}{D}}\\&\approx-\frac{1}{2}\delta r \Omega_{\rm k},
    \label{eq:linearized_velocity_physical}
\end{aligned}
\end{equation}
}
where $\Omega_{\rm k} = \sqrt{\frac{G\msmbh}{D^{3}}}$ is the Keplerian frequency about the SMBH. The sphere of influence of the BBH is always limited by $\rH$, and at this radius the linearized velocity deviates only marginally from the true velocity when $q\ll1$. We define the `wind-capture radius', $\rw$, as the radius at which a wind with the linearized velocity profile $v(\delta r)$ is marginally bound to the BBH,
\begin{equation}
    \rw \equiv \frac{2GM}{v(\rw)^2} = 2Dq^{1/3}
    \label{eq:rw_definition}
\end{equation}
This quantity differs from the Hill radius only by a constant, i.e. $\rw = 24^{1/3} \rH$. \nko{We opt to sometimes use $\rw$ instead of $\rH$ because we are borrowing intuition from Bondi-Hoyle Lyttleton accretion \citep[e.g., ][]{edgar_2004}, where the `accretion radius' describes the region within which a supersonic gas can be gravitationally captured and is defined analagously to our wind-capture radius}. Using Equation \ref{eq:linearized_velocity_physical} with $\delta r = \rw$, we define the Mach number of the wind at the wind-capture radius as, 
\begin{equation}
    \machw \equiv \frac{\Omega_{\rm k} \rw}{2\cs}.
    \label{eq:machw_definition}
\end{equation}
Using the relation $H/R = \cs/v_{\rm k}$ \citep{pringle_1981}, where $H/R$ is the disk aspect ratio, we can rewrite this expression as,
\begin{equation}
    \machw = q^{1/3}(H/R)^{-1}.
    \label{eq:machw_qHR}
\end{equation}
We have found that our accretion flow is most strongly dictated by $\machw$. This is illustrated in Figure \ref{fig:cartoon}, where we delineate three flow regimes defined by $\machw$. When $\machw \ll 1$, then $\rb \ll \rH$, and the flow accretes in a quasi-spherical, Bondi-like fashion. When $\machw > 1$, then the wind flowing through the BBH's sphere of influence is supersonic, and has high specific angular momentum. Here, the flow is best characterized as being dominated by the ram pressure of the incident wind. In the intermediate case, such that $\machw\sim 0.1-1$, the flow still has a large amount of angular momentum, but the higher gas pressure allows the wind to thermalize and produce disk-like structures. 

\subsection{Defining properties of the accretion flow}
It's worth taking a moment to  contextualize our wind tunnel framework in the landscape of other accretion flows. Our setup is similar to other local simulations of accretion flows, which have proven effective tools for understanding both single and binary BH accretion in a variety of environments, including the ISM \citep[][Schr{\o}der et al. 2022 in prep]{andrea_2019, me_2019, me_2020} and common envelope evolution\citep{macleod_2015,2015ApJ...798L..19M,2017ApJ...838...56M,ari_2017,de_2020,rosa_2020}. This family of accretion flows essentially considers permutations of Bondi-Hoyle-Lyttleton accretion \citep[for a review, see][]{edgar_2004} characterized by the gas properties of the surrounding medium. 

The case of the AGN disk is distinguished from these other scenarios by the large angular momentum content available to the embedded binary. In the limiting case of an effectively infinite angular momentum reservoir, one might expect accretion to resemble the commonly studied thin circumbinary disk \citep['CBD', ][]{munoz_2019, munoz_2020}, where BBHs are expected to expand rather than contract due to positive gravitational torques from the surrounding gas. This is similar to the scenario explored by \cite{baruteau_2011} and more recently by \cite{li_2021}, who performed global simulations of an AGN disk with an embedded binary system in two dimensions. In \cite{li_2021}, they found that prograde embedded binaries also expand, as in the case of CBDs. We emphasize, however, that for these works to globally resolve the AGN disk and the binary, they are required to use large values of $q$ and model the system in two dimensions. \nko{In works such as \cite{li_lai_2022a}, who perform two-dimensional shearing box simulations and thus can achieve higher resolutions on the scale of the CBDs and circumsingle disks, they indeed find that the flow behaves differently than in traditional CBDs. In particular, they find that the accretion flow is quasi-steady, exhibiting variability at twice the orbital frequency in the co-rotating frame. They also find that their binaries contract. However, in the two-dimensional shearing box simulations of \cite{li_2022a}, they find that the question of binary contraction versus expansion depends on the thermodynamics of the disk, and in the three-dimensional shearing box simulations of \cite{dempsey_2022}, they find that that the semimajor axis of the binary is also important.}

By using Equation \ref{eq:rw_definition}, we can write $R_{\rm w}/H\approx  2q^{1/3}(H/R)^{-1} = 2\machw$. When $q$ is large and $H/R$ is small, $R_{\rm w} \gg H$, which justifies two-dimensional approaches. However, as we will find in the following section, for much of the $q-H/R$ parameter space, $H \gtrsim R_{\rm w}$ and the resulting flow is inherently three-dimensional, supported partially by rotation and partially by pressure gradients. This suggests a dichotomy in the geometry of accretion flows surrounding embedded binaries: in thin accretion disks hosted by lower mass ($\lesssim10^6-10^7\,M_\odot$) SMBHs, the embedded binary may be better represented by the more traditional two-dimensional accretion flows studied by \cite{baruteau_2011} and \cite{li_2021}. In either thicker AGN disks (i.e., for high Eddington ratios, or in the outer Toomre-stabilized regions discussed in the next section) or for more massive ($\gtrsim10^7\,M_\odot$) SMBHs, the embedded binaries may be better represented by the vertically stratified accretion flows studied here. 

\subsection{Astrophysical context}
\label{subsec:astro_environs}
What value of $\machw$ appropriately represents the flow surrounding an embedded binary? We can answer this by applying our $\machw$ formalism to physical models of AGN disks. In particular, we consider Shakura-Sunyaev (S-S) \citep{ss1973} and Sirko-Goodman (S-G) \citep{sirko_goodman_2003} accretion disks. In the classical S-S disk, each disk annulus has Keplerian orbital energy that is gradually dissipated by some unspecified source of viscosity that is parameterized by the quantity $\alpha$. While the S-S disk forms the basis for most modern disk models, it becomes gravitationally unstable at large distances (typically $\gtrsim 10^2-10^3$ SMBH gravitational radii). The S-G model addresses these issues by including an unspecified pressure term that forces the disk to be marginally Toomre stable where it would otherwise be unstable. In reality, the disk likely continuously forms stars in this outer region, and the stellar feedback from these populations holds the disk in its marginally stable state.

\begin{figure*}
    \includegraphics[angle=270,width=\textwidth]{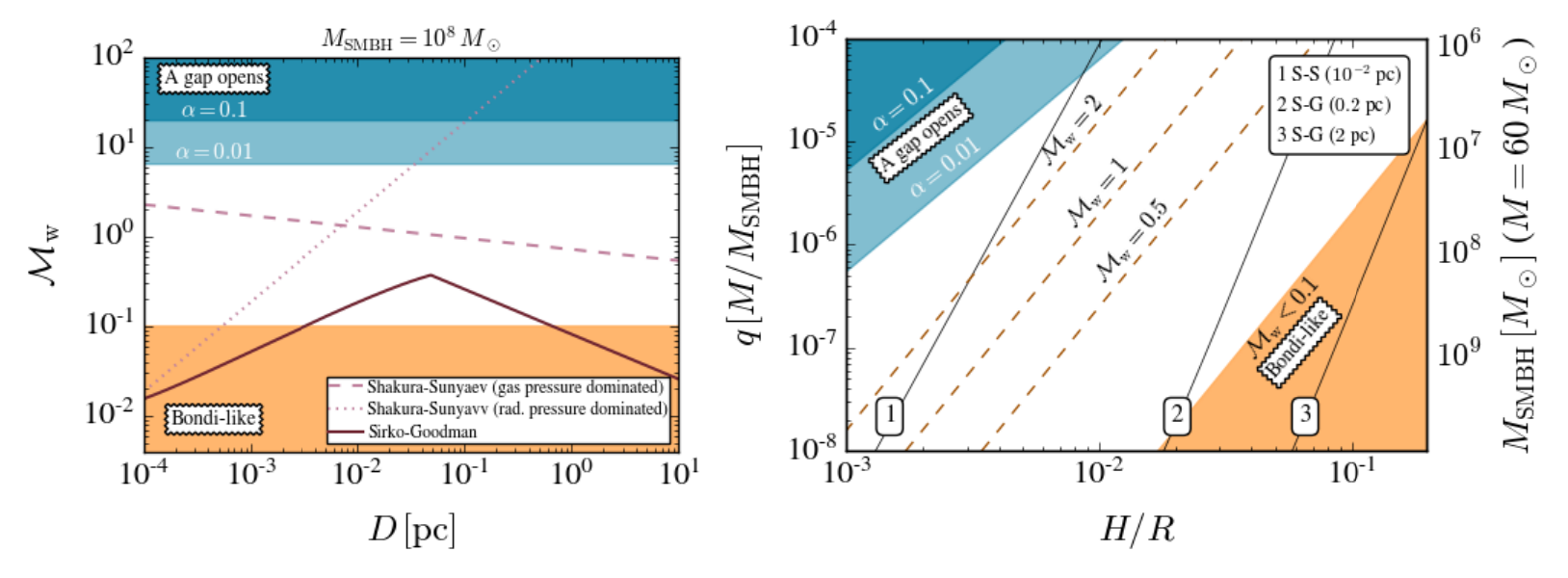}
    \caption{In both panels, we consider the dependence of $\mathcal{M}_{\rm w}$ in Shakura-Sunyaev (S-S) and Sirko-Goodman (S-G) disk models. We assume an Eddington ratio of $0.5$ and a viscosity parameter $\alpha = 0.01$. In the blue regions, we highlight where a gap is expected to open, given by the criterion in Equation 15 of \cite{crida_2006}. The orange region shows when the accretion flow becomes Bondi-like (defined such that the analytic accretion rate is $\gtrsim99\%$ of the Bondi rate). \textbf{Left panel.} Here, we plot $\mathcal{M}_{\rm w}$ as a function of the BBH distance from the central AGN for different AGN disk models. \textbf{Right panel.} We show the mass ratio ($q$) - aspect ratio ($H/R$) plane, and depict $\mathcal{M}_{\rm w}$ contours (Eq. \ref{eq:machw_qHR}) that correspond to our simulations. On the right vertical axis, we express $q$ in terms of $M_{\rm SMBH}$, given that $M=60\,M_\odot$. We also give examples of $q-H/R$ contours for different distances in both S-S and S-G models.}
    \label{fig:agn_machw}
\end{figure*}

We can explore how $\machw$ depends on S-S and S-G AGN disk models by using Equation \ref{eq:machw_qHR} to relate $\machw$ to the mass ratio $q$ and disk aspect ratio $H/R$. The profile of $\machw$ can depend sensitively on the parameters of the AGN. For simplicity, unless said otherwise, we will assume the host AGN disk has the following parameters for the remainder of this work,
\begin{itemize}
    \item A $10^8\,M_\odot$ SMBH accretes at an Eddington ratio of $\lambda_{\rm Edd}^{(\rm SMBH)} = 0.5$ with a radiative efficiency of $\eta=0.1$
    \item Viscosity is proportional to the total (gas + radiation) pressure and $\alpha=0.01$
    \item The embedded, equal-mass binary has a total mass $M=60\,M_\odot$
\end{itemize}
In the case of the S-G model, this is the same disk model used to study the orbital migration of embedded black holes in \cite{secunda_2019}, who found that their migrating BHs became trapped at roughly $\approx 10^{-3}-10^{-2}\,{\rm pc}$. 

In the left panel of Figure \ref{fig:agn_machw}, we plot $\machw$ as a function of distance $D$ in the AGN disk for both profiles. For the S-S profile, we show both a disk with $H/R$ set by gas pressure (solid line) and by radiation pressure (dashed line) \citep{accretion_power}. We also shade two regions, 
\begin{enumerate}
    \item The lower orange region which we designate as `Bondi-like' ($\machw\lesssim0.1$). Within this region, $\dot{M}$ (Eq. \ref{eq:mdot_integral}) deviates from the Bondi accretion rate \citep{bondi_1952} by only $\lesssim1\%$. If this is the case, the velocity shear in the wind is mostly negligible, and the flow geometry is quasi-spherical. 
    \item The upper blue region where our binary opens a gap within the AGN disk. This is defined by the gap opening criterion provided in Equation 15 of \cite{crida_2006}, and while it is insensitive to the specific disk profile it does depend on $\alpha$. 
\end{enumerate}
We notice a few features immediately: first, it is unusual for $\machw$ to exceed unity. An exception is the gas pressure dominated S-S curve, but only at small radii where the disk is in fact radiation pressure dominated at this value of $\lambda_{\rm Edd}^{(\rm SMBH)}$. The radiation pressure dominated S-S curve has large $\machw$ at large radii, but here the flow is gas pressure dominated. For the $S-G$ profile, which is more credible in the outer regions where the S-S disk is Toomre unstable, the flow is actually Bondi-like at regions beyond $\approx 1\,{\rm pc}$. We note the presence of the inflection point at $\approx 0.05\,{\rm pc}$ in the S-G profile, which marks the transition between the inner disk and the Toomre-stabilized outer disk \citep[see for reference Fig. 2 of ][which depicts the same profiles used here]{sirko_goodman_2003}. 

In the right panel of Figure \ref{fig:agn_machw}, we plot contours of $\machw$ in the $q-H/R$ plane (Eq. \ref{eq:machw_qHR}), and shade the same `Bondi-like' (orange) and 'gap-opening' (blue) regions as in the left panel. The vertical axis in this panel is alternatively labeled in terms of $\msmbh$, given $M=60\,M_\odot$. We have labeled $\machw=0.5,\,1$ and $2$ isocontours, which correspond to the hydrodynamic simulations that we present in Section \ref{sec:results}. The gap opening criterion is satisfied only for particularly thin accretion disks with particularly high mass ratios. This point deserves emphasis, as there are conflicting claims in the literature regarding gap formation in AGN disks. In \cite{baruteau_2011}, gap formation happens primarily because of their choice of $\alpha=8\times10^{-4}$ \citep[$\alpha\approx0.01-0.3$ is more realistic for AGN disks, e.g.][]{king_2007}, and gap opening is clearly sensitive to this parameter as can be seen in Fig. \ref{fig:agn_machw}. In \cite{bartos_2017}, they also invoke gap opening, but use a disk model that is thinner than both S-G and S-S models, causing $\machw$ to be larger. In \cite{stone_2017}, they also find that gaps are unlikely to open anywhere in the disk, and they use the AGN disk model from \cite{murray_2005}. This model is comparable to the S-G model because they also construct profiles that are marginally Toomre stable in the outer regions. Still, how significantly gap-opening affects the orbital evolution of embedded binaries remains unclear. For instance, in \cite{li_2021}, they find that the gap depth weakly affects the evolution of the semi-major axis, particularly at larger $H/R$. This is because the binary evolution is mainly dictated by gas within the Hill sphere(s), which can be readily replenished even when a gap is opened.

\section{Hydrodynamics}
\label{sec:results}

\subsection{Simulations}
\label{subsec:simulations}
To test our $\machw$ formalism, we performed three-dimensional hydrodynamic simulations of accretion onto embedded BBHs using version 4.5 of the grid-based, adaptive mesh refinement hydrodynamics code \verb|FLASH| \citep{fryxell_2000}. We use a local, wind tunnel computational domain that is continuously replenished with a shearing gaseous wind (see the arrows in the inset panels of Fig. \ref{fig:cartoon}). \nko{Physically, the velocity profile of the wind is an approximation derived from the Keplerian velocity about the SMBH in the non-inertial rest frame of the BBH, which we acquire using the linearized velocity profile defined in Equation \ref{eq:linearized_velocity_physical}}. For simplicity, we neglect gradients in the gas density and sound speed of the wind, setting them to ambient values $\rhoinf$ and $\cinf$ respectively. We evolve our simulations in a scale-free fashion, where $\cinf = \rb = 1$ and $\rhoinf=10$. We relate the linearized velocity profile in Equation \ref{eq:linearized_velocity_physical} to our scale-free setup by rewriting it as,
\begin{equation}
    v(y) = y\frac{\cinf\machw^3}{2\rb}
    \label{eq:code_velocity_profile}
\end{equation}
where we have substituted $\delta r$ for $y$. 

In this fiducial work, we solve the equations of inviscid hydrodynamics rather than the shearing box equations, and leave the latter for a later work. This means that we neglect centrifugal and coriolis accelerations from the SMBH potential. Our initial conditions in the absence of the BBH are setup to be in steady-state without these forces. When constructing the domain, we also neglect vertical gradients in the wind, acknowledging that when $\machw>1$ the AGN disk scale height $H$ becomes comparable to the BBH sphere of influence. \nko{We also neglect the breaking of the symmetric velocity profile by pressure gradients in the AGN disk. These pressure gradients however modify the velocity profile on the order of $(H/R)^2$ \citep{pringle_1981}, which is negligible for much of the parameter space (e.g., Fig. \ref{fig:agn_machw}). }

These neglected effects are negligible when $a \ll \rH$ (\nko{where $a$ is the semimajor axis of the binary}) but would affect our results when $a \gtrsim \rH$. \nko{In a real system, binaries with $a > \rH$ are unbound. Some of our simulation parameter choices correspond to this regime, but remain valuable for establishing hydrodynamic scaling relations within our wind tunnel setup.} We also neglect the self-gravity of the gas; however, even if self-gravity is important for the equilibrium state of the Toomre-unstable regions of an AGN disk, within our domain the gravity of the binary dominates and the self-gravity of the gas is dynamically unimportant. The width of our computational domain is $32\,\rb$ in each Cartesian direction, with the wind oriented in the $\pm \hat{x}$ direction and generated every timestep at the $x$ boundaries of the domain. For the $y$ and $z$ directions, we use outflow boundary conditions. \nko{Our simulation domain size is much larger than the BBH sphere of influence, and consequentially the overall gas mass in the simulation domain remains roughly constant.}

It's well-known that hierarchical triple systems are sometimes unstable \citep{eggleton_kiseleva_1995,mardling_aarseth_2001,he_petrovich_2018}. The results of \cite{eggleton_kiseleva_1995} suggest that for circular orbits, the stability of the inner binary is dictated by the semimajor axes of the inner ($a$) and outer ($D$) orbits and the mass ratio ($q$) of the inner binary to the outer massive object (the SMBH, in our case). So, while the hydrodynamics of our simulations are self-similar, they are not necessarily dynamically stable for all parameter choices. We acknowledge this as a limitation of our local simulations, but emphasize that the purpose of this work is to isolate the hydrodynamics of accreting, embedded BBHs. This allows us to develop scaling relations that can then be applied to the dynamically-stable parameter space. 

We evolve the BBHs using the active sink particles module that was developed for \verb|FLASH| by \cite{federrath_2010}, and use the same implementation of it as \cite{andrea_2019}. While we refer the reader to these works for complete descriptions of our sink prescription, the most salient details are as follows. The binary is initialized with its center of mass at $x=y=z=0$ and occupies a circular orbit with semi-major axis $a$ in the the $x-y$ plane. The $+z$ axis is parallel to the angular momentum vectors of the binary orbit and the AGN disk. We model the BBHs using Newtonian gravitational potentials with an absorbing boundary condition centered at the location of each binary companion. These absorbing boundaries have sink radii $R_{\rm s} = 0.0125\,\rb$, and any gas that crosses this boundary is removed from the domain, with the mass and angular momentum of the removed gas being recorded and added to the corresponding binary member. We also check the convergence of our mass accretion and inspiral rates with sink radius in Section \ref{app:convergence}. While the rate of inspiral is converged, the mass accretion rate decreases with decreasing sink radius. The non-convergence of the mass accretion rate with sink radius is a known issue with the sink prescription in Bondi-like accretion flows \citep{xu_2019,de_2020} and so the accretion rates derived from these simulations should be taken as upper limits. We adaptively refined around the region closest to the binary members, with smallest cell size $\delta r_{\rm min} \approx 0.0039\,\rb$. This means that there are $R_{\rm s}/\delta r_{\rm min} = 3.2$ cells across a sink radius, a value that was tested for convergence in  \cite{andrea_2019}. 

In total, we present six simulations, varying $\machw$ and $a/R_{\rm b}$, where $a$ is the BBH semimajor axis. We use values $\machw = 0.5,\,1,$ and $2$, and $a/R_{\rm b} = 0.1$ and $1$. We use a gamma-law equation of state, and assuming that the gas in the AGN disk can cool efficiently, take $\gamma = 1.1$ \footnote{In cases where the accretion rate is super-Eddington, the flow is radiation dominated and $\gamma = 4/3$ would be more appropriate, but we leave the explicit effects of radiation to be considered in a later work.}. In each simulation, the BBH occupies a circular, equatorial orbit that is prograde with the angular momentum of the wind. Each simulation was run until $t=50\,R_{\rm b}/c_\infty$, well into steady state as shown in Appendix \ref{app:convergence}. In the following two subsections, we present the results of these simulations. 

\begin{figure*}
     \includegraphics[width=0.9\textwidth]{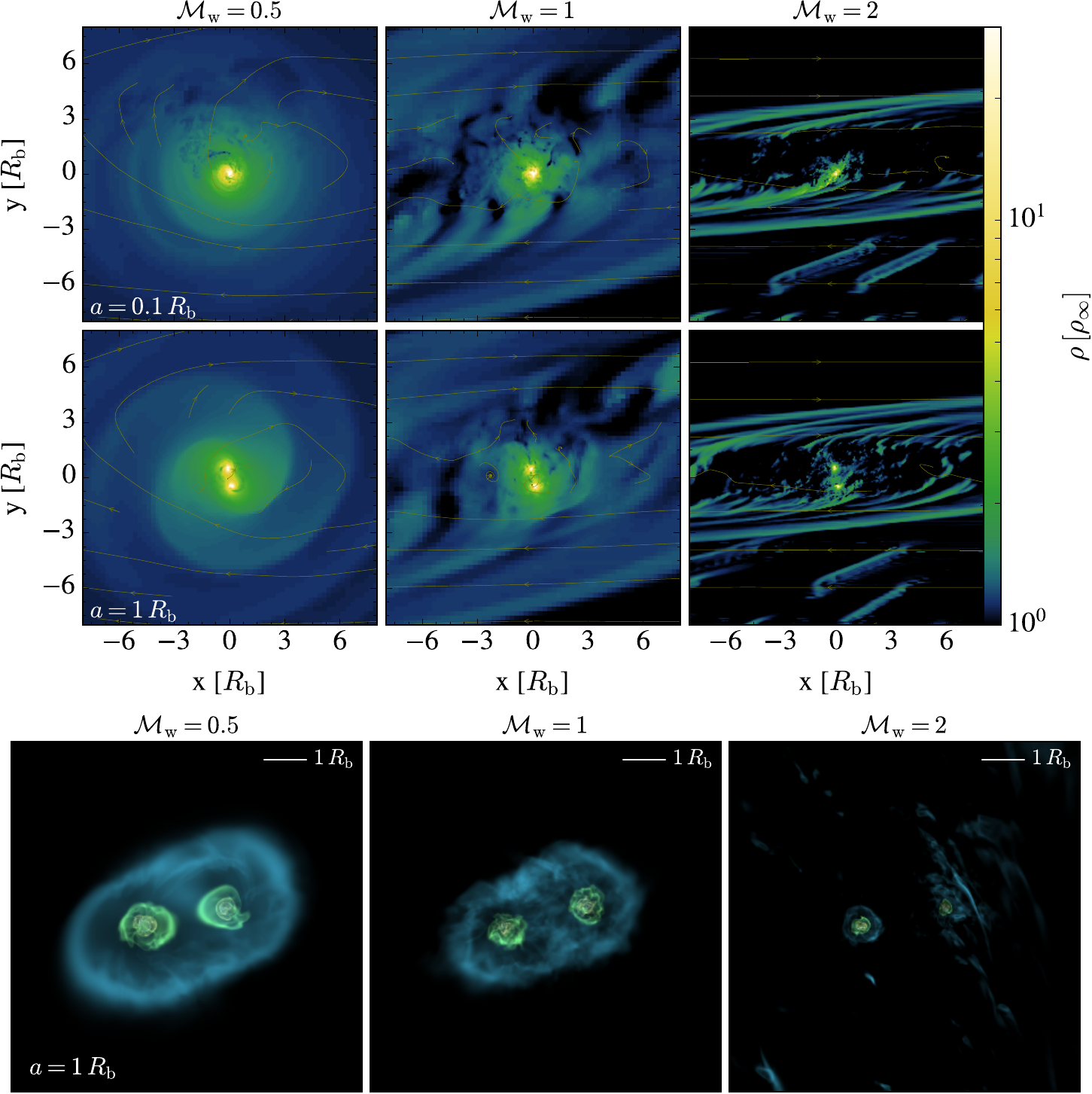}
        \caption{\textbf{Upper panel.} Here, we highlight the large scale flow morphology for each of our simulations by plotting $12\times12\,R_{\rm b}$ slices of gas density at late times. Each column from left to right represents increasing local Mach number $\machw$, where at high $\machw$ the ram pressure of the wind prevents stable flow structures from forming. In each row, we plot different values of the binary semimajor axis $a$. At small $a$, the binary shares a density enhancement, whereas at large $a$ each binary member has its own density enhancement. \textbf{Lower panel.} We plot 3D gas density isocontours for each of our $a =1\,R_{\rm b}$ simulations at the same times as the upper panel. As $\mathcal{M}_{\rm w}$ increases, the density enhancements surrounding each BH become smaller. }
\label{fig:largescale}
\end{figure*}

\subsection{Flow}
\label{subsec:flow}
In Figure \ref{fig:largescale}, we show the large-scale flow morphology of each simulation. In the top, $2\times3$ panel subfigure, we depict $12\times12\,\rb$ equatorial slices of the gas density. When $\machw=0.5$, the flow is structured and laminar, with an extended density profile that is wound in a direction prograde with the binary angular momentum. When $\machw=2$, the flow becomes `unwound', with tails that follow the direction of the wind. When $\machw=1$, the flow shares characteristics of both the $\machw=0.5$ and $2$ simulations. It's worth taking a moment to compare these structures to those found in the global simulations of \cite{li_2021} and \cite{baruteau_2011}. In these works, the disk is cool and two-dimensional, and spirals around the mini-disks of the binaries are formed and become `unwound' as they transition into the global AGN disk. While our simulations share some of these features, particularly in the bottom left and middle panels of Fig. \ref{fig:largescale}, we generally find much less pronounced spiral patterns. We hypothesize that this is due to the following distinguishing features in our accretion flow. First, the disks in our simulations are hot and partially pressure-supported, making the angular velocity profile sub-Keplerian. This, in turn, will change the radial spacing of the Lindblad resonances, which is responsible for the formation of spiral density waves. Secondly, as seen most prominently at $\machw\geq1$, our flow is has a clumpy, turbulent density distribution, which will damp any spiral density waves that are formed. Disk-like structures are evident in our 2D slices, and can be quantified with the circularization radius of the wind, 
\begin{equation}
    R_{\rm circ} = \frac{GM}{v^2(y= R_{\rm circ})} \rightarrow R_{\rm circ} = \machw^{-2}4^{1/3}\rb
    \label{eq:rcirc}
\end{equation}
For $\machw=0.5$, $1$ and $2$, the corresponding circularization radii are $R_{\rm circ}\approx6.4$, $1.6$ and $0.4\,\rb$. These values roughly correspond to the extent of the density structures in each simulation, particularly when $a/\rb = 0.1$. 

The main difference between the $a/\rb = 0.1$ and $1$ simulations is that in the former, the accretion flow forms a high-density envelope shared by both binary companions, while in the latter each companion has its own envelope. This makes sense, because when $a/\rb = 0.1$, the semimajor axis of the binary is much smaller than the sphere of influence, while when $a/\rb = 1$, they are comparable. 

In the bottom, $1\times3$ panel subfigure of Figure \ref{fig:largescale}, we show 3D gas density contours from each $a/\rb = 1$ simulation. As also seen in the 2D subfigure, when $a/\rb=1$ each binary companion has its own high-density envelope, which overlap more when $\machw$ is smaller. As $\machw$ increases, the peak density remains similar, but each density envelope becomes truncated. While in the 2D subfigure the flow appears disk-like, we see from the 3D panel that the density envelopes are as extended in the vertical directions as they are in the equatorial directions. This suggests that the flow structures surrounding embedded BBHs are partially Bondi-like, and partially disk-like. 

\subsection{Accretion \& Inspiral}
\label{subsec:inspiral}

The orbits of our simulated BBHs evolve due to a combination of gas drag and the accretion of mass and angular momentum. To compare to our numerical results, we will begin by writing down the analytic binary inspiral rate, $\dot{a}/a$. The full expression for $\dot{a}/a$, is
\begin{equation}
    \frac{\dot{a}}{a} = 2\left(\frac{\dot{L}}{L}\right)_{\rm accretion} - \frac{3\dot{M}}{M} + \left(\frac{\dot{a}}{a}\right)_{\rm drag}
    \label{eq:adot_a_all}
\end{equation}
\nko{which assumes that the specific angular momentum of the binary is Keplerian $\left(\ell = \sqrt{GMa}\right)$. We have split the angular momentum evolution into two terms for accretion and gas drag, i.e., $2\dot{L}/L = 2\left(\dot{L}/L\right)_{\rm accretion} + 2\left(\dot{L}/L\right)_{\rm drag}$, where we have written 2$\left(\dot{L}/L\right)_{\rm drag}$ as $\left(\frac{\dot{a}}{a}\right)_{\rm drag} $ to be consistent with our analytic estimate of drag which we will introduce shortly. The contribution from gas drag is} found to always be negative in our work. This is in contrast to hydrodynamic simulations of thin, gap-forming circumbinary disks where the drag term is often found to be positive, e.g. \cite{munoz_2019}. We are mainly interested in the latter two terms for mass accretion rate and drag. \nko{Note, however, we do include a time series for Eq. \ref{eq:adot_a_all} in Fig. \ref{fig:time_Convergence}.} To determine $\dot{M}$ analytically, we integrate over streamlines entering the gravitational sphere of influence of the BBH. We start by defining the length-scale,
\begin{equation}
    \rbw = \frac{\rb}{1 + \frac{1}{2}\machw^2}
    \label{eq:rbw}
\end{equation}
which smoothly transitions between $\rb$ when $\machw\ll1$ and $\rw$ when $\machw\gg1$. This allows us to continuously characterize the sphere of influence of the BBH across flow regimes analytically. The integral for the mass accretion rate is, 

\begin{equation}
    \begin{aligned}
    \dot{M}&= f\times\rhoinf \int_{r<\rbw}v(y)dA\\&= f\times4\pi\rbw^2\rhoinf\cinf\left(\frac{\machw^3}{8+4\machw^2}+1\right),
    \label{eq:mdot_integral}
    \end{aligned}
\end{equation}
where we have included an ad hoc numerical prefactor $f$. We do this because these calculations assume that \textit{all} streamlines entering the BBH sphere of influence are accreted, though in reality some fraction of this material is advected away due to gas and ram pressure, such that $f<1$. To estimate $\left(\frac{\dot{a}}{a}\right)_{\rm drag}$, we follow the approach of \cite{andrea_2019}, who determined the inspiral rate for a BBH embedded in a Bondi-Hoyle wind tunnel \citep[e.g., ][]{edgar_2004,blondin_2012}. In Appendix \ref{app:drag}, we modify their derivation for our wind tunnel set up and derive the expression

\begin{equation}
    \left(\frac{\dot{a}}{a}\right)_{\rm drag} = s\times\frac{8\pi G a \rho_\infty}{(GM/4a + \cinf^2 + \machw^2\cinf^2)^{1/2}}\left(1+\frac{2R_{\rm bw}}{a}\right)^p
\label{eq:adot_a_drag}
\end{equation}

Here, $s$ and $p$ are numerically determined parameters that we will fit for. This calculation assumes that the orbital energy of the binary is dissipated by a gravitational wake induced by each component of the binary as they orbit through a dense gaseous envelope. These wake-induced gravitational torques are different than those that dictate the orbital angular momentum evolution in thin circumbinary disks, which is mainly driven by Lindblad torques from the inner and outer disks. Our main argument for neglecting the study of Lindblad torques is that our accretion flows are pressure dominated, rather than rotation dominated, and we don't have any circumbinary disks. It is possible that there is some transition, that occurs at higher $\machw$, where the angular momentum evolution of the binary transitions to a structure that better resembles the usual circumbinary disk. 

\begin{figure}[bht]
        \includegraphics[width=\textwidth]{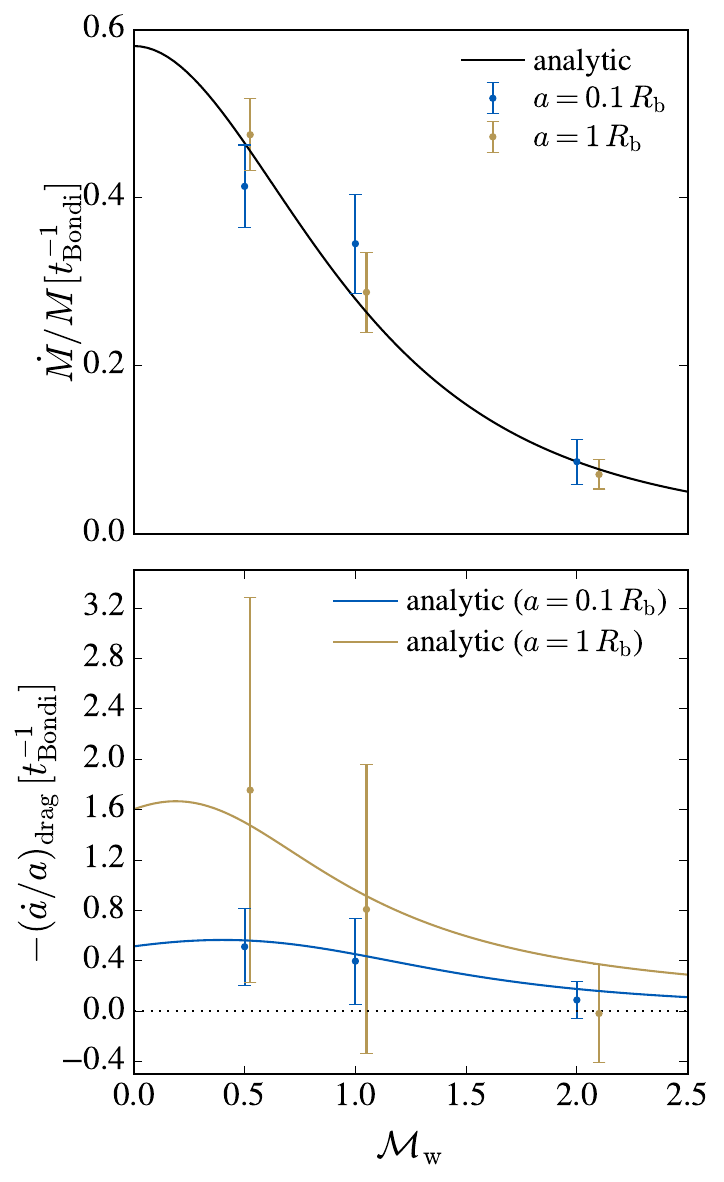}
    \caption{We compare our simulated accretion (top panel) and drag (bottom panel) rates to our analytic estimates (Equations \ref{eq:adot_a_all}-\ref{eq:adot_a_drag}) as a function of $\machw$. Both quantities are expressed in units of $t_{\rm Bondi}^{-1}$, which is the Bondi accretion rate per unit mass ($= \cinf^3/4\pi G^2M\rhoinf $). As $\machw\rightarrow0$, both profiles approach a Bondi solution. Care should be taken in extrapolating to $\machw\gg1$, as at high values $\rw \gg H$ and the flow geometry becomes more two-dimensional, making our wind tunnel simulations less appropriate \citep[see][for comparison]{baruteau_2011}.
    }
    \label{fig:inspiral}
\end{figure}

In Figure \ref{fig:inspiral}, we assess how well our analytic estimates compare to our simulated values for the accretion and drag rates as a function of $\machw$. Our simulated value of mass accretion rate are calculated by taking the time derivative of the mass of the binary, which is updated each time-step during our simulation. The drag rate is determined by recording the net drag force that the gas exerts on the binary at each timestep. We express our rates in units of $t_{\rm Bondi}^{-1}= \cinf^3/4\pi G^2M\rhoinf$, which is the Bondi accretion rate per unit mass \citep[][]{bondi_1952}. We have numerically determined that $f=0.58$ (Eq. \ref{eq:mdot_integral}), $s = 0.35$ and $p=-0.86$ (Eq. \ref{eq:adot_a_drag}). Using these adjustments, Fig. \ref{fig:inspiral} shows that our analytic estimates for our mass accretion and drag rates reasonably reproduce our simulated results. In general, the binaries accrete and inspiral on similar timescales, with inspiral occurring faster (slower) at larger (smaller) separations. This is in contrast to what is seen in Bondi-Hoyle accretion onto binaries, e.g. Fig. 11 of \cite{andrea_2019}, where inspiral happens about $3-4$ times faster than accretion. We note that when $\machw\gg1$, our local wind tunnel approximation breaks down because the BBH can begin influencing the AGN disk, so care should be taken in extrapolating to larger values of $\machw$. In those cases, the flow geometry is likely better represented by the results of \cite{baruteau_2011}. In contrast, extrapolating our mass accretion prescription to $\machw=0$ provides no issue, as the flow approaches a Bondi solution in that limit. 

\section{Evolution of embedded BBHs}
\label{sec:evolution}

\subsection{How does the binary first become embedded?}
\label{subsec:binary_ic}
To determine the evolution of embedded BBHs, it is important to understand how they are first captured. The BBH is either born in the outer regions of the disk or, having previously existed in the nuclear cluster when the disk was formed, it gradually aligns with the disk. The initial position of the embedded BBH is determined by how it became embedded, which strongly impacts the binary's subsequent evolution. 

Roughly $\approx 80\%$ of stars in the nuclear cluster are expected to have formed in situ \citep{antonini_2015}. If the BBH is formed directly in the disk \citep[as studied in ][]{stone_2017}, it by definition will start in the Toomre-stabilized region as this is where star formation occurs. In Fig. 5 of \cite{murray_2005}, they generally find that the star formation rate (SFR) in nuclear starburst disks increases as a function of radius. This would suggest that the vast majority of embedded BBHs formed in situ will begin their journeys at large ($\gtrsim1-10\,{\rm pc}$) radii in the AGN disk. The possibility that they merge in the inner regions of the AGN disk depends on the relative efficiency of inspiral versus migration. 

The dynamical capture of BBHs by AGN disks depends on the distribution of stellar-mass BHs (sBHs) in the nuclear cluster (NC). This distribution is often chosen by invoking the inferred distribution of the black hole cusp surrounding Sag A*, which roughly scales as $\propto D^{-2.5}$ \citep{bartko_2009,alexander_hopman_2009} and likely has $\sim1000$ sBHs in the inner $0.1\,{\rm pc}$ \citep{antonini_2014,hailey_2018}. However, the formation of this cusp requires that the stellar population has relaxed around the central SMBH \citep{BW76}. The relaxation timescale for a $10^6\,M_\odot$ SMBH can range from $0.1-10\,{\rm Gyr}$ \citep{oleary_2009}, and can exceed a Hubble time for more massive SMBHs. So, it may be that heavier SMBHs harbor no black hole cusp during the AGN phase, which makes the viability of the dynamical capture channel less certain. For this reason, we favor the in situ formation of embedded BBHs, which should begin their journey in the AGN disk at larger ($\gtrsim1\,{\rm pc}$) distances. 

\subsection{Binary Evolution}
\label{subsec:binary_evo}
Once BBHs become embedded in the AGN disk, they will simultaneously inspiral, migrate and grow in mass. In this section, we model these processes by evolving a system of coupled ordinary differential equations for $\dot{D}$ (migration), $\dot{M}$ (accretion), and $\dot{a}$ (inspiral). To guide our intuition, we begin by estimating the timescales associated with each of these processes. We use an S-G AGN disk model with $M_{\rm SMBH}=10^8$, $\alpha=0.01$, $\lambda_{\rm Edd}^{(\rm SMBH)} = 0.5$ and assume that the viscosity is proportional to the total (gas + radiation) pressure. At a characteristic distance of $1\,{\rm pc}$ for a $60\, M_\odot$ binary, the migration timescale is
\begin{equation}
    \begin{aligned}
    t_{\rm migr} \approx 61\,{\rm Myr}&\left(\frac{0.06}{H/R}\right)^{-2}\left(\frac{60\,M_\odot}{M_{\rm BBH}}\right)^{-1}\left(\frac{10^8\,M_\odot}{M_{\rm SMBH}}\right)^{3/2}\\ \times &\left(\frac{0.5\,{\rm pc}}{D}\right)^{-1/2}\left(\frac{\Sigma}{1540\,{\rm g\, cm^{-2}}}\right)^{-1}
    \end{aligned}
    \label{eq:timescale_migration}
\end{equation}
which we acquire from Equation 10 of \cite{paardekooper_2014}. This characteristic timescale is comparable lifetime of the host disk ($\approx 50-150\,{\rm Myr})$, suggesting that embedded binaries may migrate significantly \citep[e,g, ][]{secunda_2019,secunda_2020}. This equation is for the static torque usually associated with Type I migration, which is valid under our assumption that the embedded BBH only negligibly affects the evolution of the host AGN disk. This expression also neglects self-gravity of the host disk and radiative torques, which may enhance the migration rate \citep{kley_nelson_2012}.

At large semimajor axes, before gravitational-wave emission dominates, binaries inspiral due to the combined influence of mass accretion and drag. At distances of $\gtrsim0.5\,{\rm pc}$, the accretion flow is Bondi-like (e.g., Fig. \ref{fig:agn_machw}), for which the mass doubling time ($\equiv M/\dot{M}_{\rm Bondi}$) is characteristically
\begin{equation}
    \begin{aligned}
    t_{\rm Bondi} \approx 148\,{\rm Kyr} &\left(\frac{M}{60\,M_\odot}\right)^{-1}\left(\frac{\rho}{8\times10^{-15}\,{\rm g\,cm^{-1}}}\right)\\&\times\left(\frac{c_s}{54\,{\rm km\,s^{-1}}}\right)^{-3}
    \end{aligned}
    \label{eq:timescale_bondi}
\end{equation}

This is extremely rapid, which at first glance suggests the embedded binaries should easily balloon in mass. The associated drag timescale ($\equiv$ the inverse of Eq. \ref{eq:adot_a_drag}, where for simplicity we take $\machw=0$ and $a=100\,{\rm au}$) is $\approx18\,{\rm Kyr}$, increasing to $\approx72\,{\rm Kyr}$ at $a=10\,{\rm au}$. Together, these three timescales form a hierarchy: the binary inspirals first (mainly due to drag, but also due to mass accretion), then increases in mass, then migrates. 

\begin{figure}
    \centering
    \includegraphics[width=\textwidth]{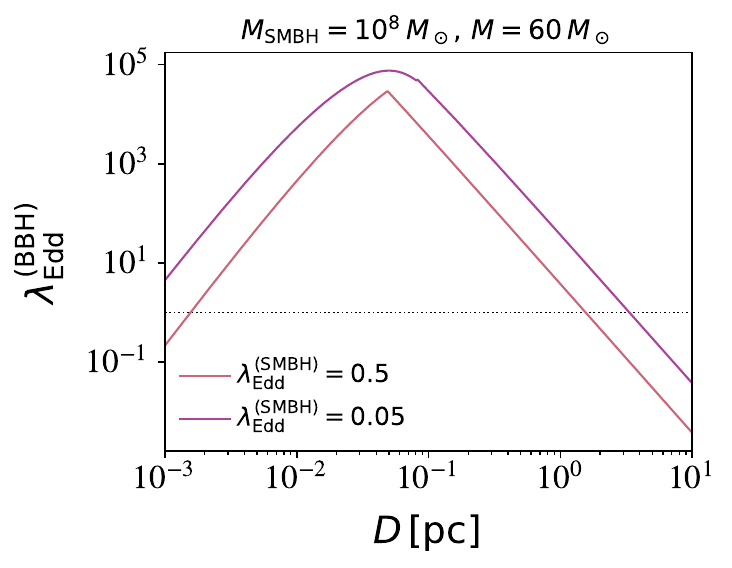}
    \caption{We plot the embedded BBH's Eddington ratio, $\lambda^{(\rm BBH)}_{\rm Edd}$, as a function of distance in the AGN disk. Here, we consider the case of a $60\,M_\odot$ binary embedded in an AGN disk with host mass $M_{\rm SMBH}=10^8\,M_\odot$ accreting at Eddington ratios $\lambda_{\rm Edd}^{(\rm SMBH)} = 0.5$  (red) and $0.05$ (purple). We assume that the embedded binary accretes with a radiative efficiency of $\eta=0.01$. For most regions in the AGN disk, the embedded binary accretes at highly super-Eddington rates.}
    \label{fig:eddratios}
\end{figure}

However, the mass accretion rate associated with Equation \ref{eq:timescale_bondi} is super-Eddington throughout much of the disk. This can be seen in Figure \ref{fig:eddratios}, where we plot the Eddington ratio of the embedded binary, $\lambda^{(\rm BBH)}_{\rm Edd}$, as a function of distance in the host AGN disk. This figure shows that the embedded binary can be supplied gas at rates of up to $\sim10^4-10^5$ the Eddington ratio. These profiles peak at radii of $\sim0.05-0.07\,{\rm pc}$, where there is a transition to the star-forming outer regions of the host AGN disk.  In practice, we limit the mass accretion rate to its Eddington-limited value, 

\begin{equation}
    \dot{M}_{\rm Edd} = \frac{L_{\rm Edd}}{\eta c^2}
\end{equation}
Simulations of super-Eddington accretion have shown that mass accretion rates can significantly exceed the Eddington rate \citep{mckinney_2014,mckinney_2015,dai_2018},
which is essentially set by how small the radiative efficiency ($\eta$) is. Thin disks have typical values of $\eta=0.1$, which lowers to roughly $\eta=0.01$ near the Eddington limit, and can be even smaller at significantly super-Eddington rates. For simplicity, we adopt a constant value of $\eta = 0.01$. The associated Eddington accretion timescale ($\equiv M/\dot{M}_{\rm Edd}$) is independent of mass,
\begin{equation}
    t_{\rm Edd} \approx 4.5\,{\rm Myr}\left(\frac{\eta}{0.01}\right)
    \label{eq:timescale_edd}
\end{equation}

Even with Eddington-limiting our mass accretion rate, it still occurs on a shorter timescale than migration. We must make a few modifications for the inspiral rate in our binary evolution prescriptions, as well. First, we include a term for the gravitational wave inspiral rate of an equal-mass binary \citep{peters_1964}. Second, we neglect the $(\dot{L}/L)_{\rm accretion}$ term in Eq. \ref{eq:adot_a_all}, since in real systems this term can only contribute marginally ($\mathcal{O}(\dot{M}l_{\rm isco})$, where $l_{\rm isco}$ is the specific angular momentum at the innermost stable circular orbit) to the binary's evolution. We also make the assumption that whenever the mass accretion rate is Eddington limited, so is the drag rate, since an Eddington-limited disk will be depleted of the gas that drives the binary to inspiral.

\begin{figure}[bht]
\includegraphics[width=\textwidth,angle=0]{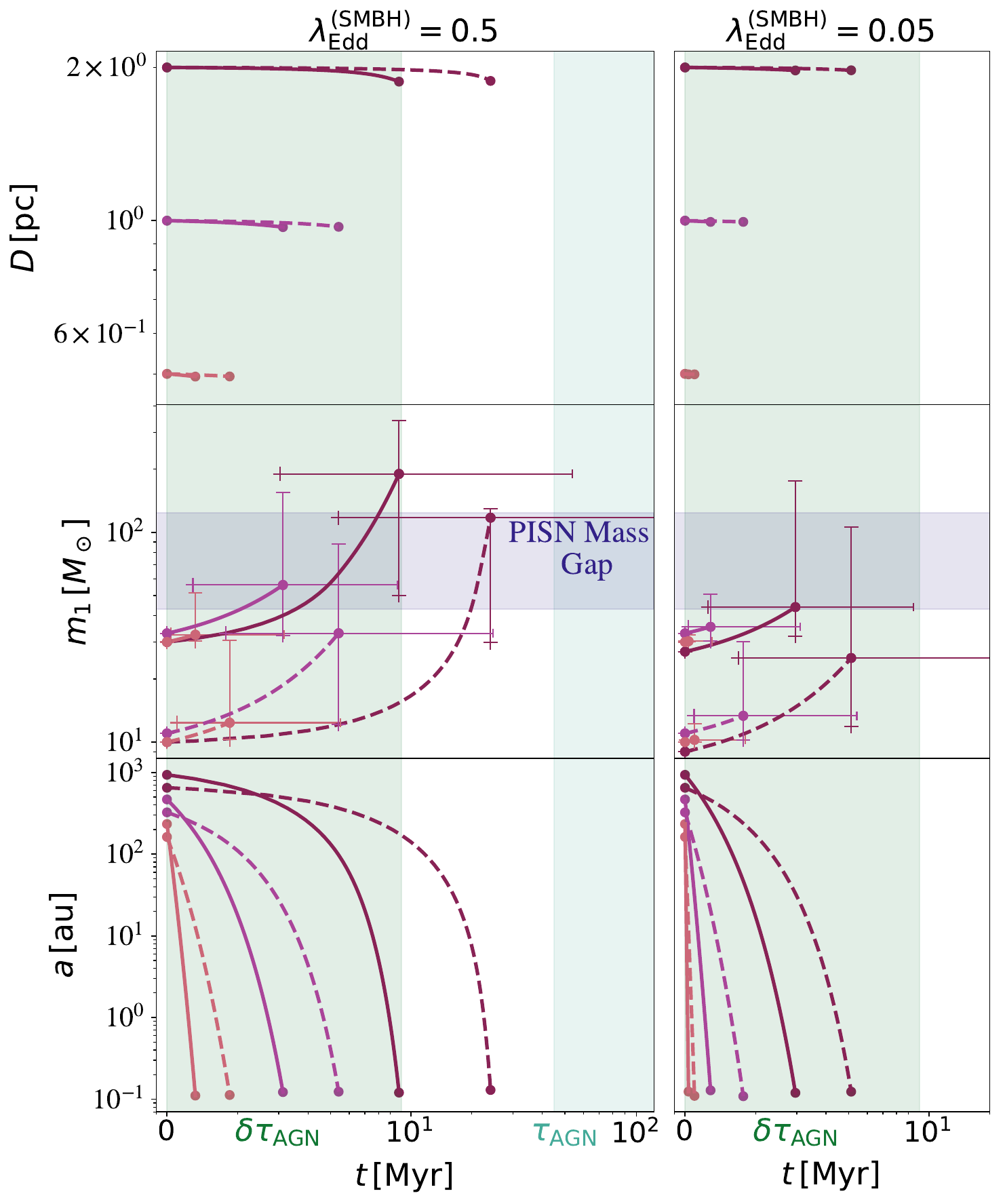}
    \caption{We plot the integrated evolution of binaries embedded in an AGN disk with initial separations $a$ from $162$ to $650\,{\rm au}$, initial distances from the SMBH $D = [0.5,\,1,\,2]\,{\rm pc}$ and initial masses $M = m_1+m_2 = 20$ \nko{(dashed lines)} and $60\, M_\odot$ \nko{(solid lines)}. The initial separations as set to be marginally stable to the \cite{eggleton_kiseleva_1995} hierarchical triple instability. We specifically consider migration within the AGN disk, mass growth, and inspiral rate. We do this for Sirko-Goodman disk models for Eddington ratios of 0.5 and 0.05, where we take $M_{\rm SMBH} = 10^8$, $\alpha=0.01$, and assume that viscosity is proportional to the total (gas+radiation) pressure. We shade regions based on the typical duty cycle of an AGN, $\tau_{\rm AGN}$, and on the timescale on which AGN disks flicker and restructure themselves, $\delta \tau_{\rm AGN}$. 
    Shaded in purple is the mass range associated with the pair instability supernova mass gap. The error bars in the middle panel are derived by rerunning the same profiles with the prescibed mass accretion rate increased or decreased by a factor of 10.}
    \label{fig:binaryEvo}
\end{figure}

We evolve our binaries in Figure \ref{fig:binaryEvo}, where we show how the binary separation, mass, and AGN disk position change in time. We initialize our evolved binaries with masses $M = m_1+m_2 = 20$ \nko{(dashed lines)} and $60\,M_\odot$ \nko{(solid lines)}, and initial distances $D = 0.5$, $1$, and $2\,{\rm pc}$. We initialize our binary semimajor axes such that they are marginally stable to the hierarchical triple instability outlined in \cite{eggleton_kiseleva_1995}.  We use $\lambda_{\rm Edd}^{(\rm SMBH)} = 0.5$ on the left subfigure and $\lambda_{\rm Edd}^{(\rm SMBH)} = 0.05$ on the right. Each integration is terminated when the semimajor axis reaches $a = 0.1\,{\rm au}$, after which the inspiral becomes rapid due to gravitational-wave decay. In all cases, the full evolution occurs on timescales of $\lesssim 1-30{\rm Myr}$. The relative hierarchy of timescales follows our expectations from our analytic predictions; the binary inspirals the quickest, and is driven to inspiral by a combination of drag, gravitational-wave radiation, and mass accretion. Second, the masses begins to increase, and in some cases double. For each profile, the binaries migrate only marginally before they merge. To highlight uncertainties in our binary evolution prescriptions, we reran each profile with the prescribed mass accretion rate reduced or increased by a factor of ten, the results of which are shown by the error bars in our middle panel. 

In general, the evolution of the binary happens well within the lifetime of the AGN disk ($\approx 50-150\,{\rm Myr}$, labeled $\tau_{\rm AGN}$ in Fig. \ref{fig:binaryEvo}). However, a potentially important caveat in this evolution is that we keep the AGN disk static. In reality, the AGN disk can `flicker' on short timescales ($\approx 0.1-10\,{\rm Myr}$, labeled $\delta\tau_{\rm AGN}$ in Fig. \ref{fig:binaryEvo}) \citep[e.g., ][]{schawinski_2015}. Similar phenomena is seen in the simulations of \cite{angles-alcazar_2020}, who found that the gas supplying luminous quasars can restructure and change orientation on timescales of $\approx 0.1-1\,{\rm Myr}$. If the binary becomes embedded, but then within $\delta\tau_{\rm AGN}$ the disk restructures and reorients itself, the BBH may once again require its inclination to be damped with respect to the restructured disk for a gas-assisted inspiral to proceed.

\section{Discussion}
\label{sec:discussion}

\subsection{Implications for gravitational-wave signals}
\label{subsec:gw}

The results of Figure \ref{fig:binaryEvo} have profound implications for the gravitational-wave signals observed by LIGO and Virgo. If a BBH becomes embedded within an AGN disk, it can be driven to merger on timescales $\lesssim1-30\,{\rm Myr}$. While the rates at which BBHs form within AGN disks is unclear, our calculations suggest that gas-assisted inspiral is a viable formation channel for BBH merger events. Furthermore, the BHs in these binaries accrete at significant rates, suggesting that a pair of $30+30\,M_\odot$ BHs can potentially evolve into the pair instability supernova mass gap. Such a system was recently observed by LIGO/Virgo \citep{GW190521a, GW190521b}, with a reported possible association with an electromagnetic transient event in an AGN \citep{graham_2020}. If further analysis confirms the association or if additional associations are identified with future LIGO events, such signals would amount to a confirmation of the merger channel studied here. 

A caveat is that the accreting BHs in our paradigm should significantly spin up and become aligned, leading to a large effective spin parameter, $\chi_{\rm eff}\gtrsim0.9$. In contrast, the GW190521 gravitational wave signal was associated with a much smaller value of $\chi_{\rm eff}=0.08$. This suggests that the BHs either had low spins individually or had large spins that were misaligned with the orbit. If GW190521 did come from the AGN disk formation channel, then there must be some difference in the evolutionary process that we characterized in Section \ref{subsec:binary_evo}. A few plausible scenarios are,
\begin{enumerate}
    \item While we focus on embedded BBHs, individual BHs will migrate and accrete mass from the disk in an analogous way. The inner parsec surrounding the host SMBH may be replete with these single BHs, each of which can potentially grow to sufficiently large masses that they enter the PISN mass gap. These BHs could then pair and merge dynamically later in their life. In this scenario the BH spins would be randomly oriented, potentially leading to a low $\chi_{\rm eff}$, regardless of the individual BH spins.
    \item The scenario presented here involves no dynamical interactions while the BBH evolves within the disk. We can calculate the interaction time between a BBH and a tertiary star with a randomly oriented orbit: 
    \begin{eqnarray*}
        t &=& \frac{1}{n \sigma V} \nonumber \\
          &\simeq& 2 \left( \frac{n}{10^3\ {\rm pc}^{-3}} \right)^{-1} \left( \frac{a}{100\ {\rm au}} \right)^{-2} \left( \frac{v_{\rm k}}{650\ {\rm km}\ {\rm s}^{-1}} \right)^{-1} {\rm Myr},
    \end{eqnarray*}
    where we have adopted a cross-section, $\sigma$, derived from the orbital separation, and an interaction velocity $V$ corresponding to the Keplerian orbit of a BBH at a distance of $\sim1$ pc from a $10^8\,M_\odot$ SMBH. For a stellar density of 10$^3$ pc$^{-3}$ and an orbital separation of 10$^2$ au, we find an interaction time of $\approx $2 Myr, shorter than a merger time of $\approx $10 Myr. These strong dynamical encounters can fundamentally alter the BBHs orbit, reorienting the orbital angular momentum vector, altering the orbital eccentricity, and even trading companions. However, from Figure \ref{fig:binaryEvo}, note that most of the mass accreted by a BH occurs when the orbital separation shrinks to $\lesssim$~10 AU, with a much longer interaction time. 
    \item The accretion flow studied here is initially laminar, while AGN disk models typically assumed some $\alpha$ viscosity that parameterizes the role of isotropic, turbulent eddies that permeate the AGN disk. The characteristic length-scale of an eddy is $\sim\alpha H$, and depending on the value of $\machw$, this can be comparable to $\rb$. If so, the evolution of the embedded binary could be driven more strongly by the accretion of eddies with randomly distributed angular momenta than by the angular momentum supplied from the AGN disk velocity profile. If this is the case, mass accretion might occur similarly, but the spin evolution of the binaries would be stochastic. This is, however, a different flow geometry than studied here and deserves attention in its own right. 
\end{enumerate}

While the relative importance of AGN disks as a formation scenario for BBH mergers may remain unresolved even after future LIGO/Virgo observing runs, the space-based gravitational wave detector LISA may be sensitive to BBHs in AGN, a prospect that we will consider now. Although many different factors can impact the overall rate of systems observed by LISA, at sufficiently small orbital separations where gravitational wave radiation dominates the orbital evolution of point masses such as BBHs, general relativity makes a clear prediction for the distribution of BBH orbital separations, $P(a)$; since the strength of gravitational wave radiation increases with decreasing orbital separation, $(\dot{a}/a)_{\rm GR} \sim a^{-4}$ \citep{peters_1964}, a population of equal-mass BBH binaries will have $P(a)\sim a^3$. However, at sufficiently large separations gas drag forces will dominate. Taking $p\approx-1$, Equation \ref{eq:adot_a_drag} indicates that $(\dot{a}/a)_{\rm drag}\propto a^{5/2}$ when $a\ll\rb$, producing an orbital separation distribution $P(a)\sim a^{-7/2}$. By equating these two forces, we can find the critical orbital separation, $a_{\rm crit}$, indicating the transition between the two regimes:
\begin{equation}
    \begin{aligned}
    a_{\rm crit} \simeq&\,25.7\left(\frac{m_1}{30\,M_\odot}\right)^{9/13}\left(c_\infty/20\,{\rm km\,s^{-1}}\right)^{-4/13}\\ &\times(\rho_\infty/10^{-11}\,{\rm g\,cm^{-3}})^{-2/13}
    \end{aligned}
    \label{eq:a_crit}
\end{equation}
This corresponds to an orbital period $\approx$ days or a gravitational wave frequency $\approx$ $\mu{\rm Hz}$. This is, unfortunately, well below the LISA sensitivity curve which extends down to $\approx$ mHz. Additionally, the strong dependence on $a$ of both the GW inspiral rate and our drag rate makes the dependence of $a_{\rm crit}$ on the gas density and sound speed extremely weak. The binaries may also grow significantly by accretion during their inspiral, which would cause $a_{\rm crit}$ to increase further. Equation \ref{eq:a_crit} also assumes a purely hydrodynamic drag rate, while in reality the accretion flow is highly super-Eddington which likely further decreases the efficiency of drag. Thus, if our $a$ scalings predicted in Eq. \ref{eq:adot_a_drag} hold, we find it unlikely for LISA to be able to detect embedded BBHs.

\subsection{Electromagnetic signatures}
\label{subsec:em}
In recent years, the possibility that BBH mergers may be accompanied by electromagnetic (EM) signatures has intrigued the astrophysics community. There have been many proposed mechanisms for producing a signature, including the rapid accretion of relic disks post-merger \citep[][Schr{\o}der et al. 2021 in prep]{perna_2016,sophie_2018}, the shock-heating of circumbinary disks due to the post-merger recoil \citep{corrales_2010, demink_2017}, and mergers following a single-star progenitor \citep{dorazio_2018}. In \cite{bartos_2017}, they also studied the fates of embedded BBHs, and argued that these binaries could produce observable signatures by reaching super-Eddington luminosities via relativistic, beamed outflows. Interest in this scenario has been reinvigorated by the claimed association of an AGN flare with BBH merger GW190521. While this claim is tenuous \citep{ashton_2020}, it highlights one of the difficulties of this scenario: even if an embedded BBH produces a luminous signal, it must be distinguished from the dominant AGN emission. \nko{We note the recent work, \cite{graham_2020}, which compared AGN flaring activity with BBH merger candidates observed by LIGO/Virgo in O3, and found 9 possible associations. These associations depend on the nature of electromagnetic signatures from merging embedded BBHs, the definition of what `typical' AGN flaring is, and on the large localization windows from gravitational-wave observations. Given these considerations, we expect that the identification of EM signatures from embedded BBH mergers will only be made robust when we better understand what an associated flare would look like.}

Radiation may be produced by an embedded BBH either by accretion during the inspiral phase or transiently immediately following the merger. We turn our attention, first, to the case of a steady-state luminosity. In many cases, the embedded BBH is supplied\footnote{We intentionally avoid the term \textit{accreted}, here, because while we can confidently estimate the gas supply rate via Equation \ref{eq:mdot_integral}, the actual fraction of accreted material on sub-grid scales is uncertain.} gas at rates near the Eddington limit or significantly above it \citep[as pointed out by other authors, e.g., ][]{stone_2017,bartos_2017}. However, even in cases where the mass accretion rate can exceed the Eddington limit, the luminosity itself will still be limited because radiation becomes trapped within the flow. In this case, the accretion flow will be a geometrically and optically thick advection-dominated accretion flow \citep['ADAF', ][]{narayan_1994}, producing an unusually soft blackbody spectrum. For an Eddington-limited pair of stellar-mass black holes to be observable, they must compete with the host AGN. The relative luminosity of the BBH accretor to the AGN is,
\begin{equation}
    \frac{L_{\rm BBH}}{L_{\rm AGN}} \sim 10^{-7}  \left(\frac{\lambda_{\rm Edd}^{\rm (SMBH)}}{0.1}\right)\left(\frac{M}{60\,M_\odot}\right)\left(\frac{M_{\rm SMBH}}{10^8\,M_\odot}\right)^{-1}
\end{equation}
For typical values of $\lambda_{\rm Edd}^{\rm(SMBH)}$, $M_{\rm BBH}$, and $M_{\rm SMBH}$, this value is extremely small and the accreting BBH will be indistinguishable from the host AGN unless the two are resolved. 

Clearly, if the electromagnetic signatures from an embedded BBH are to be observed, it must shine brighter than the Eddington luminosity. The most natural way for this to happen is by producing a collimated, relativistic jet that has its emission beamed towards us \citep[e.g.,][]{2014ApJ...794....9M}. We can estimate an upper limit on the jet luminosity by assuming the flow is magnetically arrested, producing powerful Blandford-Znajek jets due to large poloidal magnetic fields \citep{blandorf,tchekhovskoy_2011}. Assuming this, we take
\begin{equation}
    L_{\rm jet} = \eta_{\rm MAD}\dot{M}c^2
\end{equation}
where $\eta_{\rm MAD}$ is the jet efficiency for a magnetically arrested disk (MAD). The jet efficiency in MADs commonly reaches $\sim200-300\%$ by tapping the rotational energy of the BH \citep{tchekhovskoy_2011}, so we assume a characteristic $\eta_{\rm MAD}=2$. Assuming that the radiative efficiency of the accretion flow is $\eta \approx 0.1$, then the total jet luminosity is
\begin{equation}
    L_{\rm jet} \approx 20\left(\frac{\eta_{\rm MAD}}{2}\right)\left(\frac{\eta}{0.1}\right)^{-1}L_{\rm Edd}
    \label{eq:jet_luminosity_physical}
\end{equation}
If this jet is directed towards us, the emission will be beamed within an opening angle $\theta \sim 1/\Gamma$, where $\Gamma$ is the Lorentz factor of the jet. If we measure the flux from this jet at earth, then the inferred isotropic luminosity is,
\begin{equation}
    L_{\rm iso} \sim \frac{4\pi}{\theta^2} \frac{L_{\rm jet}}{2} = \Gamma^2 2\pi L_{\rm jet}
\end{equation}
If we compare this boosted luminosity to the host AGN luminosity, then we find that
\begin{equation}
    \begin{aligned}
    \frac{L_{\rm iso}}{L_{\rm AGN}} \sim &10^{-5}\Gamma^2  \left(\frac{2}{\eta_{\rm MAD}}\right)\left(\frac{0.1}{\eta}\right)^{-1}\left(\frac{\lambda_{\rm Edd}^{\rm (SMBH)}}{0.1}\right)\\&\times\left(\frac{M}{60\,M_\odot}\right)\left(\frac{M_{\rm SMBH}}{10^8\,M_\odot}\right)^{-1}
    \end{aligned}
\end{equation}
If we require that $L_{\rm iso} \gtrsim L_{\rm AGN}$, then the corresponding $\Gamma$ for which this is achieved is $\Gamma\approx316$. This is in the upper range of $\Gamma$ values associated with gamma-ray bursts \citep{lithwick_2001, gehrels_2007}, 
and corresponds to an opening angle $\theta\sim0.18^\circ$. Even if embedded BBHs could produce jets with this Lorentz factor, we would only see $\times \theta^2/4 \sim 0.0002\%$ of them due to the inclination-dependence of the emission. Furthermore, while \cite{bartos_2017} suggest that embedded BBHs form gaps, allowing jets to escape unimpeded, Fig. \ref{fig:agn_machw} suggests most binaries will not form gaps, and Fig. \ref{fig:binaryEvo} suggests most binaries merge in the outer regions ($\gtrsim0.5-1\,{\rm pc}$) of the disk. In these regions, $\machw\sim0.1$, and $H$ is roughly two orders of magnitude larger than the binary sphere of influence. This provides a large column density of material for the jet to traverse, likely resulting in significant mass-loading which would lower the resulting $\Gamma$. This suggests that it is difficult to detect a jet associated with an embedded BBH unless it occurs at frequencies separate from the dominant AGN emission. We note that the EM signatures associated with these jets have been studied in detail by two recent papers; \cite{zhu_2021}, for the case of embedded neutron stars, and \cite{perna_2021}, for jet propagation from both BHs and neutron stars. In both works, they emphasize the issue of distinguishing the transient emission from the AGN emission, and find that the transient can more easily outshine the AGN in the X-ray, infrared and radio bands. 

Finally, we consider the possibility of transient EM signatures that occur during or immediately following merger. There are two possibilities; the first possibility is that the binary experiences a transient spike in its accretion rate and thus luminosity post-merger \citep{milosavljevic_2005}. However, these studies were focused on thin circumbinary disks with carved out cavities, and in the thick flows that accompany these binaries no such spike is likely to exist. We therefore ignore this possibility. The other possibility is that the sudden decrease in gravitational potential at the onset of merger rapidly shocks the surrounding gaseous envelope, producing intense photospheric emission. 

The sudden mass loss accompanying the BBH merger is associated with a drop of potential energy in the surrounding high-density envelope,
\begin{equation}
    \Delta E = \int_{R_{\rm bw}}\frac{G\Delta M_{\rm merger}}{r} \rho(r)dV
\end{equation}
We assume that within the BBH sphere of influence, characterized by $R_{\rm bw}$ (Eq. \ref{eq:rbw}), the density profile is Bondi-like, i.e. $\rho(r) \sim \rhoinf(r/R_{\rm bw})^{-3/2}$ \citep{shapiro_teukolsky_1983}. This yields a change in potential energy, 
\begin{equation}
    \Delta E = 4\pi\rhoinf G\Delta M_{\rm merger}R_{\rm bw}^2
\end{equation}
If we assume the flow is Bondi-like in the region of interest and that roughly $\sim50\%$ of the BBH mass is lost upon merger, then this change in energy can be rewritten as,
\begin{equation}
    \begin{aligned}
    \Delta E \approx 8.1\times10^{40}\,{\rm ergs}&\left(\frac{M}{60\,M_\odot}\right)^3\left(\frac{\rhoinf}{10^{-18}\,{\rm g\,cm^{-3}}}\right)\\&\times\left(\frac{\cinf}{25\,{\rm km\,s^{-1}}}\right)^{-4}
    \end{aligned}
\end{equation}
If we crudely assume that this energy is lost within a sound-crossing time $\Delta t = \rb/\cinf$, then the luminosity produced by the shock ($\approx \Delta E/\Delta t$) is
\begin{equation}
    \begin{aligned}
    L_{\rm shock} \approx &1.6\times10^{32}\,{\rm ergs\,s^{-1}}&\left(\frac{M}{60\,M_\odot}\right)^2\left(\frac{\rhoinf}{10^{-18}\,{\rm g\,cm^{-3}}}\right)\\&\times\left(\frac{\cinf}{25\,{\rm km\,s^{-1}}}\right)^{-1}
    \end{aligned}
    \label{eq:luminosity_shock}
\end{equation}
Since a near-Eddington, $10^8\,M_\odot$ host SMBH should have a luminosity of roughly $L_{\rm AGN}\approx10^{45}-10^{46}\,{\rm ergs\,s^{-1}}$, it is implausible for the luminosity of the post-merger shock to exceed that of the AGN. 

\subsection{Feedback}
One of the uncertainties in the accretion flow embedding the binaries is the role of feedback. Feedback can be deposited into the gas supply either radiatively or mechanically, through winds and relativistic jets, and can effect the mass accretion rate and efficiency of drag. 

If an accreting binary produces a jet, it is possible for the jet to carve out a cavity and suppress the gas supply \citep[e.g.][]{2002MNRAS.337.1349R}. If the force exerted on the ambient medium by a relativistic jet $\sim L_{\rm jet}/c$, then it will exert a ram pressure $\sim L_{\rm jet}/ c \theta^2 r^2$, where $\theta$ is the opening angle of the jet. By assuming that as the jet plunges through the medium, it becomes subrelativistic and spreads laterally, we can take $\theta^2=4\pi$. We can set the jet ram pressure equal to the gas pressure of the ambient medium to determine the radius of the jet-medium interface (denoted $r_j$), 
\begin{equation}
    \begin{aligned}
    & \frac{L_{\rm jet}}{4\pi r_j^2c} = \rhoinf\cinf^2, \\
    & r_j = \sqrt{\frac{L_{\rm jet}}{4\pi\rhoinf\cinf^2 c}}
    \end{aligned}
\end{equation}
To compare this to the relevant physical scale, we can write $r_j$ in terms of the Bondi radius and physical values,
\begin{equation}
    \begin{aligned}
    r_j/R_{\rm b} &\approx 56\left(\frac{\lambda_{\rm Edd}^{(\rm jet)}}{20}\right)^{1/2}\left(\frac{M}{60\,M_\odot}\right)^{-1/2}\\
    &\times\left(\frac{\rhoinf}{10^{-18}\,{\rm g\,cm^{-3}}}\right)^{-1/2}\left(\frac{\cinf}{25\,{\rm km\,s^{-1}}}\right)
    \end{aligned}
\end{equation}
In this relation, we've expressed the jet luminosity in terms of its Eddington ratio, $\lambda_{\rm Edd}^{(\rm jet)}$. For the assumed ambient conditions (typical of the AGN disk at $\approx 1\,{\rm pc}$) and Eddington ratio (an upper limit, see Eq. \ref{eq:jet_luminosity_physical}), $r_j$ is much larger than $R_{\rm b}$. This suggests that if a powerful jet is produced by the binary, it should easily carve out a large cavity and impede gas supply. This could at least transiently diminish the accretion rate and may impose a duty cycle on the growth of the BBH. Alternatively, we can use $r_j < R_{\rm B}$ as the condition required for a cavity not to form. Then, the maximum Eddington ratio of the jet is,
\begin{equation}
    \lambda_{\rm Edd}^{(\rm jet)} < 0.0064\left(\frac{M}{60\,M_\odot}\right)
    \left(\frac{\rhoinf}{10^{-18}\,{\rm g\,cm^{-3}}}\right)\left(\frac{\cinf}{25\,{\rm km\,s^{-1}}}\right)^{-2}
\end{equation}
This is an extremely low jet efficiency and would likely require a black hole with a very small spin \citep{narayan_2003}. However, closer to $\approx 10^{-1}-10^{-2}\,{\rm pc}$, the gas density can reach $\approx 10^{-14}\,{\rm g\,cm^{-3}}$ \citep{sirko_goodman_2003}, which would result in a minimum Eddington ratio of $64$. This suggests that the ability of a jet to impede the gas supply strongly depends on where the BBH is in the disk; at larger distances, it likely induces a duty cycle in the accretion flow, while at smaller distances the jet will be quickly extinguished. 

\subsection{Caveats}
\label{subsec:caveats}
The results of this work, particularly the evolutionary tracks laid out in Section \ref{subsec:binary_evo}, have uncertainties that deserve emphasis. Here are a few essential ingredients required to properly understand the evolution of embedded BBHs:
\begin{itemize}
    \item Understanding how drag and mass accretion rates change in the super-Eddington regime is critical to understanding the evolution of embedded BBHs. The ratio of the accretion to inspiral timescale is what determines the final mass of the binary upon merger and will be altered if radiation is realistically taken into account. Additionally, the radiative efficiency affects both the mass accretion rate and the evolutionary timescale. While we use a constant radiative efficiency of $\eta=0.01$, it will be higher in the sub-Eddington regime, but can be even lower in the highly super-Eddington regime. While the details of the evolutionary tracks depend on how drag in the super-Eddington regime is prescribed, two main features are clear: that embedded BBHs can grow significantly, possibly entering the PISN mass gap before merging, and that the time-to-merger is $\lesssim1-30\,{\rm Myr}$. 
    \item The true nature of AGN disks, particularly in the outer Toomre-unstable region, remains an open question. While we favor the profiles of \cite{sirko_goodman_2003}, they use an unspecified pressure source to stabilize this region. This provides a useful model for making predictions but remains a crude estimate which should be explored further. Additionally, as discussed in Section \ref{subsec:binary_evo}, the time-dependence of AGN disks is a major uncertainty. As found in the simulations of \cite{angles-alcazar_2020}, at parsec scales the disk can reorient itself on $\approx {\rm Myr}$ timescales, which may alter the occupation fraction of black holes in the AGN disk or limit the total amount of time the binaries have to evolve in the disk. 
    \item We assumed that the embedded BBH is allowed to evolve through the AGN undisturbed, but in reality it dynamically couples to the surrounding stellar population. Three-body scattering may merge the BBH faster \citep{stone_2017} or may reorient the binaries with respect to the surrounding flow, which can alter the mass and spin evolution of the binaries prior to merger. Additionally, as discussed in Section \ref{subsec:binary_ic}, a proper understanding of the black hole distribution near the SMBH will allow us to better understand how many BHs can be captured dynamically by the AGN and where their initial position in the disk is, which can strongly affect their subsequent evolution. 
    \nko{\item We have neglected multiple effects in our hydrodynamic model. Importantly, these include the non-inertial forces of coriolis and centrifugal forces, which will likely modify the flow. Additionally, properly modelling the vertical stratification of the flow is also likely to be important. Even given a more realistic domain, there are large uncertainties pertaining both to the size of the sink radius (which will affect the convergence of the mass accretion rate (e.g. Section \ref{app:convergence}) and on the neglected physics. In particular, a proper treatment of the thermodynamics (i.e., via a cooling prescription or better yet radiation transport) will affect the mass accretion and drag rates \citep[e.g., ][]{li_2022a}, and feedback processes (such as relativistic jets or outflows launched on subgrid scales) may also alter the flow substantially.}
\end{itemize}

\subsection{Summary}
\label{sec:events}
To summarize, we describe the chronology of embedded BBHs as they evolve hydrodynamically through the host AGN disk:
\begin{itemize}
    \item \textit{Birth.} A pair of black holes are either born in the disk (typically $\gtrsim0.1-1\,{\rm pc}$), or they are dynamically captured. If the nuclear cluster harbors a black hole cusp, it is possible for the black holes to be initially captured at much smaller radii. As discussed in Section \ref{subsec:binary_ic}, it is unclear whether or not a black hole cusp will have been formed during the AGN phase of the SMBH, particularly for higher mass SMBHs where the relaxation timescale of the surrounding stellar population can be longer than the age of the Universe. For this reason, we find it more likely that BBHs first become embedded in the outer regions of the AGN disk. 
    \item \textit{Accretion.} As an embedded BBH orbits prograde with the AGN disk, it will become engulfed in a gaseous wind that has an asymmetric velocity profile in the rest frame of the BBH. This is depicted in Figures \ref{fig:cartoon}, \ref{fig:agn_machw} and \ref{fig:largescale}, where the characteristics of the accretion flow are primarily defined by $\machw$, the Mach number of the wind in the BBH rest frame. At large $\machw$, the ram pressure of the wind dominates the accretion flow, and at small $\machw$ the accretion flow becomes Bondi-like. In the outer, Toomre unstable regions of an AGN disk, $\machw$ is often small enough that Bondi accretion provides an adequate approximation (Fig. \ref{fig:agn_machw}). When $\machw$ is large enough ($\gg 1$), the accretion flow becomes planar, likely resembling a pair of thin mini-disks and potentially opening an annular gap in the AGN disk \citep{baruteau_2011,li_2021}. Our analysis is invalid in this regime, which likely holds for AGN disks that are geometrically thin and are hosted by lighter ($\sim10^6\,M_\odot$) SMBHs. This makes our results most salient for binaries embedded within the disks of heavier ($\gtrsim10^7\,M_\odot$) SMBHs. For most regions in the disk, a BBH will accrete at super-Eddington rates (Section \ref{subsec:binary_evo}); if the BBH accretes at the Eddington limit for $\gtrsim5\,{\rm Myr}$, it can double its mass.
    \item \textit{Inspiral.} The inspiral of the BBH is most strongly governed by mass accretion and drag, with drag generally contributing more than mass accretion (Fig. \ref{fig:inspiral}). In Section \ref{subsec:inspiral}, we provide analytic expressions with best-fit numerical coefficients that reasonably reproduce the inspiral rate measured in our hydrodynamic simulations. We find that typical BBHs, beginning their journey anywhere between $0.1$ and $2\,{\rm pc}$ in the AGN disk, will merge within $\lesssim 1-30\,{\rm Myr}$. 
    \item \textit{Migration.} Concurrent to accretion and inspiral, embedded BBHs will migrate in the AGN disk (typically inwards). For the initial distances we considered ($0.5-2\,{\rm pc}$), we found that our BBHs migrated only marginally before merging.
    
    \item \textit{Multimessenger signatures.} In Section \ref{sec:evolution}, we found that BBHs can often double their mass during their inspiral. This can alter the resulting black hole mass spectrum detected by LIGO, allowing embedded BBHs to enter the pair-instability supernova mass gap, as discussed in Section \ref{subsec:gw}. In Section \ref{subsec:em}, we studied the possible EM signatures accompanying an accreting BBH or occurring transiently post-merger. In general, it is difficult for the BBH to outshine its host AGN. If the BBH produced a relativistic jet, it would have to have a large Lorentz factor to be detectable, leaving only a very small fraction of these sources detectable due to the inclination-dependence of the beamed emission. Shock heating in the AGN disk due to the post-merger recoil can produce electromagnetic transients, but with luminosities far less than that of the AGN.

\end{itemize}

\section*{Acknowledgements}
The notions expressed in this work have grown out of several exchanges with K. Auchettl, D. D'Orazio, Z. Haiman, K. Kremer, D. Lin, J. Samsing, and A. Tchekhovskoy. We are indebted to them for guidance and encouragement. We acknowledge support from the  Heising-Simons Foundation, the Danish National Research Foundation (DNRF132) and NSF (AST-1911206 and AST-1852393), and the Vera Rubin Presidential Chair for Diversity at UCSC. The authors thank the Niels Bohr Institute for its hospitality while part of this work was completed. J.J.A.\ acknowledges funding from CIERA through a Postdoctoral Fellowship. A.A. is grateful for support from the Berkeley and Cranor Fellowships at U.C. Berkeley and the NSF Graduate Research Fellowship program (Grant No. DGE 1752814). The simulations presented in this work were run on the HPC facility at the University of Copenhagen, funded by a grant from VILLUM FONDEN (project number 16599), and the lux supercomputer at UC Santa Cruz, funded by NSF MRI grant AST 1828315. The simulations presented in this work were performed using software produced by the Flash Center for Computational Science at the University of Chicago, which was visualized using code managed by the yt Project.

\appendix
\section{Simulation Convergence}
\label{app:convergence}

\begin{figure}
\includegraphics[width=0.85\textwidth]{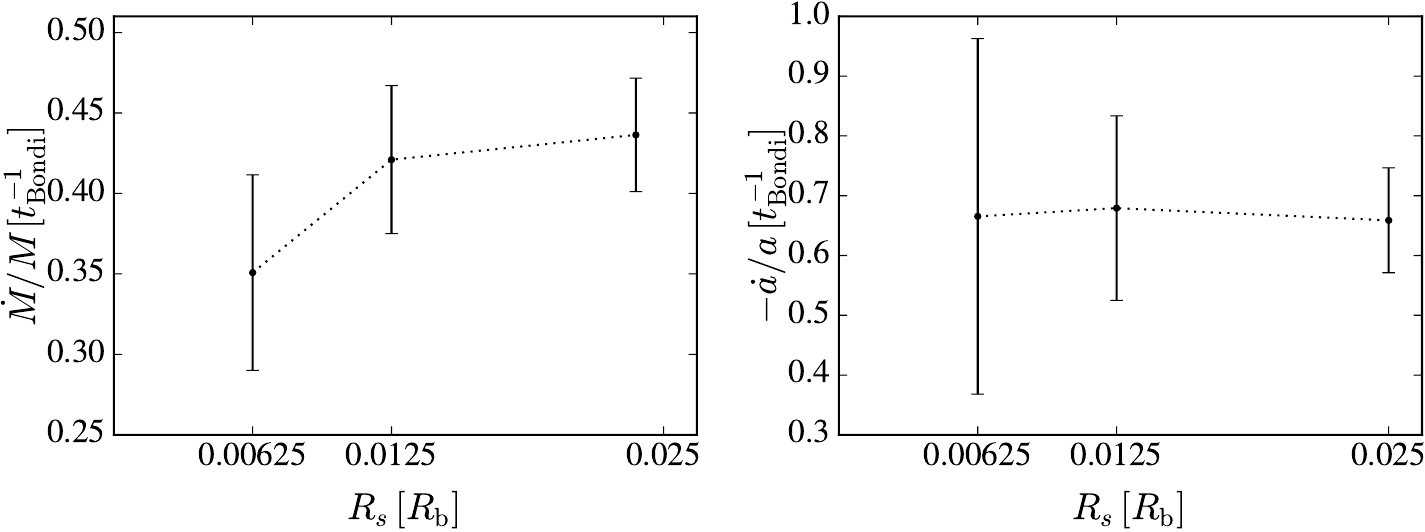}
    \caption{We plot the time-averaged accretion and inspiral rates for different sink radii in our $a=0.1\,R_{\rm b},\,\machw=0.5$ simulation. $R_{\rm s} = 0.0125\,R_{\rm b}$ was the value standardly used in our simulations. $R_{\rm s} = 0.00625\,R_{\rm b}$ is computationally expensive, as it requires increasing the resolution to resolve the boundary adequately, so this simulation was restarted from the $R_{\rm s} = 0.0125\,R_{\rm b}$ run at $t=40\,R_{\rm b}/c_\infty$ and ran for $10$ more accretion timescales. For $R_{\rm s}=0.025\,R_{\rm b}$, we ran both a full simulation and one restarted at $t=40\,R_{\rm b}/c_\infty$, and both resulted in the same average accretion and inspiral rates.}
    \label{fig:sink_convergence}
\end{figure}

 In our simulations, we represented the boundaries of the black holes with an absorbing boundary condition of radius $R_{\rm s}=0.0125\,\rb$. In Figure \ref{fig:sink_convergence}, we show the convergence of our accretion and inspiral rates as a function of $R_{\rm s}$ by comparing them at $R_{\rm s} = 0.0625,\,0.0125$ and $0.025\,\rb$. A full simulation run at $R_{\rm s}=0.0625\,\rb$ is computationally expensive, so we opted to restart the $R_{\rm s} = 0.0125\,\rb$ simulation at $t = 40\,\rb/\cinf$ (well into steady state) and run it for $10$ accretion timescales. For $R_{\rm s} = 0.0625\,\rb$, we also increased the maximum AMR level so the number of cells across a sink radius would remain constant. We used this restart approach for $R_{\rm s} = 0.025\,\rb$ as well, and also compared it to running the full simulation duration at that sink radius, and found no difference in the result. In the specific case of Fig. \ref{fig:sink_convergence}, we used the simulation with $a = 0.1\,\rb$ and $\machw=0.5$. We find that that the accretion rate typically decreases with sink radius and that the inspiral rate is relatively constant albeit with higher variability at smaller sink radii. This is expected; as the sink radius decreases, more streamlines are deflected before reaching the boundary, decreasing the accretion rate \citep[e.g., ][]{xu_2019}. On the other hand, gravitational drag results from the cumulative influence of all gas in the binary's sphere of influence, and so is less sensitive to the sink radius. This suggests that our results regarding drag in Section \ref{subsec:inspiral} should be relatively robust, but the mass accretion rate should be taken as an upper limit.

\begin{figure}
\includegraphics[width=0.65\textwidth]{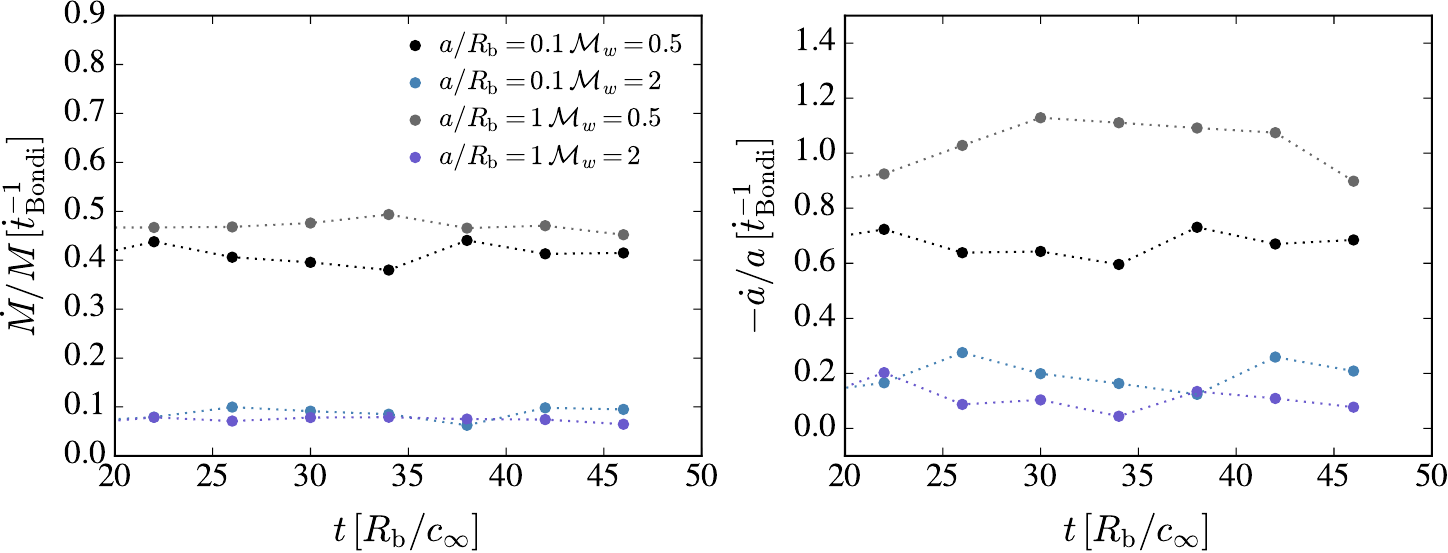}
    \caption{We plot the accretion and inspiral rates for several of our simulations for the majority of the simulation run-time. Since these rates have strong variability, we averaged both quantities in $4\,R_{\rm b}/c_{\infty}$ bins, represented by each scatter point.
    The plots depict times $>20\, R_{\rm b}/c_\infty$ after which the flow is in quasi-steady state.}
    \label{fig:time_Convergence}
\end{figure}

In Figure \ref{fig:time_Convergence}, we plot the accretion and inspiral rates for the majority of the simulation run-time for each simulation (excluding the intermediate $\machw=1$ simulations). \nko{Here, both mass and inspiral rates are evolved on the fly in our simulation, and account for all terms in Eq. \ref{eq:adot_a_drag}.} In general, both rates are highly variable, so we opted to plot averages over both quantities in $t = 4\,R_{\rm b}/c_{\infty}$ bins. We depict both rates at times $>20\,R_{\rm b}/c_\infty$, after transient spikes in the accretion and inspiral rates have died off and the flow reaches steady state. 

\section{Derivation of Drag Prescription}
\label{app:drag}
Here, we derive the drag formula given by Equation \ref{eq:adot_a_drag}, where we follow analogously from Appendix A of \cite{andrea_2019}. As a binary companion of mass $m_1$ orbits, it will capture gas into a dense posterior wake, to which we associate a characteristic radius,
\begin{equation}
    R_{\rm BH,1} = \frac{2Gm_1}{V_1^2 + \cinf^2 + \machw^2 \cinf^2}
\end{equation}
This is essentially the Bondi-Hoyle accretion radius, with terms in the denominator for the orbital velocity of the binary companion ($V_1$), the sound speed ($\cinf$), and the characteristic velocity of the shearing wind $\machw\cinf$. We can relate this to a characteristic energy dissipation rate, which is defined as the kinetic energy flux entering the accretion cross-section $\pi R_{\rm BH,1}^2$,
\begin{equation}
    \dot{E}_{\rm BH,1} = \frac{1}{2}\pi R_{\rm BH,1}^2\rho\left(V_1^2 + \cinf^2 + \machw^2 \cinf^2\right)^{3/2} 
\end{equation}
Since we are considering an equal-mass binary, the total energy dissipation rate of the binary is double this. We also take $V_1 = v_{\rm orb}/2 = \sqrt{GM/a}/2$ where $M=m_1+m_2$ is the total mass of the binary,
\begin{equation}
\dot{E}_{\rm BH,orb} = \pi R_{\rm BH,1}^2\rho\left(GM/4a + \cinf^2 + \machw\cinf^2\right)^{3/2} = \pi G^2M^2\rho\left(GM/4a + \cinf^2 + \machw^2 \cinf^2\right)^{-1/2}
\end{equation}
We also need to make a choice for the gas density near the binary orbit. In spherically symmetric, Bondi accretion, the density scales as $\rho(r) = \rhoinf(r/R_{\rm b})^{-3/2}$ at radii $r\ll R_{\rm b}$. In our case, the density profile will be altered due to the presence of rotation. Additionally, we want a form of the density profile that asymptotically approaches $\rhoinf$ when the separation of the binary system is very large. Instead of normalizing the radial profile to $R_{\rm b}$, we instead normalize it to $R_{\rm bw}$ (see Eq. \ref{eq:rbw}). Under these assumptions we define our density profile as $\rho(r)=\left(1+\frac{R_{\rm bw}}{r}\right)^p$, where $p$ is to be determined from our simulation results. We can then write our dissipation rate as, 
\begin{equation}
    \dot{E}_{\rm BH,orb} = \pi G^2M^2\rho_\infty\left(V_1^2 + \cinf^2 + \machw^2 \cinf^2\right)^{-1/2}\left(1 + \frac{2R_{\rm bw}}{a}\right)^p
\end{equation}
where we have evaluated the density profile at the binary position $r=a/2$. Finally, we divide this quantity by the orbital energy $-E_{\rm BH,orb} = -GM^2/8a$ to acquire an inspiral rate, and also add a prefactor $s$ to scale our results,
\begin{equation}
    \left(\frac{\dot{a}}{a}\right)_{\rm drag} = s\times 8\pi Ga\rho_\infty\left(V_1^2 + \cinf^2 + \machw \cinf^2\right)^{-1/2}\left(1 + \frac{2R_{\rm bw}}{a}\right)^p
\end{equation}
The prefactor is necessary because our energy dissipation rate assumes that streamlines enter the cross-section(s) $\pi R_{\rm BH,1}^2$ ballistically (i.e., unhindered by pressure support) and because we use simple characteristic estimates for our velocities.  Still, we expect the overall scaling of this expression to hold.

\bibliographystyle{aasjournal}
\bibliography{references}

\end{document}